\documentclass[pra,twocolumn,superscriptaddress,secnumarabic,balancelastpage,amsmath,amssymb]{revtex4-1}
\usepackage{amssymb}
\usepackage{amsmath}
\usepackage[dvipsnames]{xcolor}
\usepackage{chngpage}
\usepackage{lgrind}        % convert program listings to a form includable in a LaTeX document
\usepackage{color}         % produces boxes or entire pages with colored backgrounds
\usepackage{graphics}      % standard graphics specifications
\usepackage[pdftex]{graphicx}      % alternative graphics specifications
\usepackage{longtable}     % helps with long table options
\usepackage{epsf}          % old package handles encapsulated post script issues
\usepackage{bm}            % special 'bold-math' package
\usepackage{asymptote}     % For typesetting of mathematical illustrations
\usepackage{thumbpdf}
\usepackage{verbatim}
\usepackage{url}
\usepackage{MnSymbol}
\usepackage{physics}
\usepackage[colorlinks=true]{hyperref} 
\usepackage{enumitem}	% compact enumeration

\usepackage{float}
\usepackage{times}
\usepackage{color}
\usepackage{verbatim}
\usepackage{soul,xcolor}
\usepackage{tipa, extraipa}
\setstcolor{red}
\usepackage{colortbl}

\newcommand{\Op}{\mathcal{O}}
\newcommand{\cc}{\sublptr{\textit{c}}}

\newcommand{\cpl}{\acute{c}}%\textsubplus{\textit{c}}
\newcommand{\cm}{\grave{c}}%\textsubbar{\textit{c}}}

\usepackage{tabularx}
\usepackage[normalem]{ulem} % For \sout (and hence \cut)
\makeatletter
\AtBeginDocument{
\g@addto@macro{\appendix}{\renewcommand{\p@subsection}{\@Alph\c@section}}
}
\makeatother

\begin{document}

\title{Higher-Order Methods for Hamiltonian Engineering Pulse Sequence Design}

\affiliation{Department of Physics, Harvard University, Cambridge, Massachusetts 02138, USA}

\author{Matthew Tyler$^{1}$}
\thanks{These authors contributed equally to this work}
\author{Hengyun Zhou$^{1}$}
\thanks{These authors contributed equally to this work}
\author{Leigh S. Martin$^{1}$}
\author{Nathaniel Leitao$^1$}
\author{Mikhail D. Lukin$^1$}
\email{lukin@physics.harvard.edu}

% Include the date command, but leave its argument blank.

\begin{abstract}
We introduce a framework for designing Hamiltonian engineering pulse sequences that systematically accounts for the effects of higher-order contributions to the Floquet-Magnus expansion.
Our techniques result in simple, intuitive decoupling rules, despite the higher-order contributions naively involving complicated, non-local-in-time commutators.
We illustrate how these rules can be used to efficiently design improved Hamiltonian engineering pulse sequences for a wide variety of tasks, such as dynamical decoupling, quantum sensing, and quantum simulation.
\end{abstract}
\maketitle

%\tableofcontents

\section{Introduction and Motivation}
The effective control of many-body quantum dynamics is an important challenge in the emerging field of quantum science and technology, with wide-ranging applications in quantum computation~\cite{ladd2010quantum,vandersypen2005nmr}, quantum sensing~\cite{degen2017quantum}, and quantum simulation~\cite{georgescu2014quantum,bloch2012quantum}.
One of the key tools for controlling such many-body quantum dynamics is Hamiltonian engineering~\cite{waugh1968approach,burum1979analysis,cory1990time,choi2020robust,viola1999dynamical,khodjasteh2005fault,uhrig2007keeping,alvarez2015localization,wei2018exploring,wei2019emergent,choi2020robust,hayes2014programmable,ajoy2013quantum,choi2017dynamical,haas2019engineering}, in which a train of pulses transform the original system Hamiltonian into a desired target Hamiltonian for various applications.
Indeed, from the inception of such techniques in early NMR work to the present day, Hamiltonian engineering has enabled high resolution spectroscopy~\cite{waugh1968approach,cory1990time,rose2018high,slichter2013principles,mehring2012principles}, high sensitivity metrology~\cite{zhou2020quantum}, as well as the realization of exotic Floquet phases of matter~\cite{choi2017observation,zhang2017observation,lindner2011floquet}.

One of the key tools for performing Hamiltonian engineering is average Hamiltonian theory~\cite{haeberlen1968coherent}.
Here, the engineered Hamiltonian is approximated by the time-average of interaction-picture Hamiltonians with respect to the control pulses.
This allows the effective engineering of many-body Hamiltonians, even in the case where only global manipulation of spins is accessible, as is the case in many large-scale quantum systems~\cite{kucsko2018critical,bloch2012quantum}.
Moreover, design rules that systematically take into account robustness against various imperfections can be derived~\cite{burum1979analysis,choi2020robust}, enabling robust pulse sequence design as well.

Despite the success of techniques based on average Hamiltonian theory, large variations in performance still exist among the different sequences obtained, suggesting that higher-order contributions in the full Magnus expansion may play an important role.
Existing works treating higher-order contributions often rely purely on symmetrization, or treat the higher-order terms on a case-by-case basis~\cite{burum1979analysis,choi2020robust,cory1990multiple}.
However, finding \textit{general} conditions for the cancellation of higher-order Magnus terms can be non-trivial, as the expressions involve commutators that are \textit{non-local} in time.

\begin{figure}
\begin{center}
\includegraphics[width=\columnwidth]{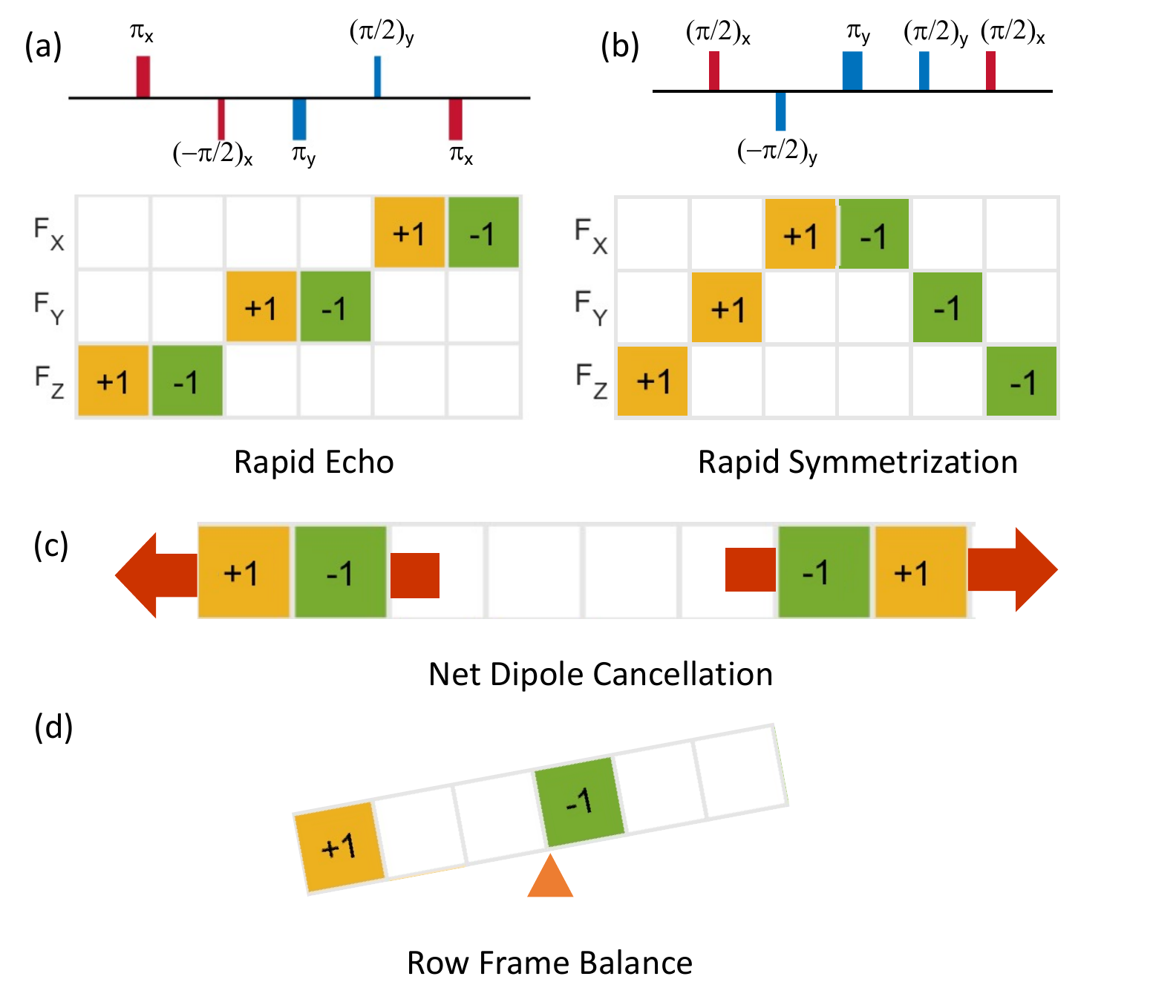}
\caption{{\bf Visualization of representative first-order cancellation rules.} (a) A rapid spin echo block, which cancels disorder-disorder and disorder-Ising terms. (b) Rapid symmetrization block, which cancels Ising-Ising terms. (c) Cancellation of net dipoles along each axis, which cancels disorder-Heisenberg terms. (d) Balancing the weight in each row, which cancels Ising-Heisenberg terms.}
\label{fig:cancellation}
\end{center}
\end{figure}

In this paper, we systematically analyze higher-order Magnus contributions to effective Hamiltonians, providing a general toolset for pulse sequence design in interacting spin systems in the form of concise decoupling rules.
Despite the non-local nature of the commutators involved in higher-order contributions, we are still able to generalize many results from average Hamiltonian theory. First, we find that the frame representation employed in Ref.~\cite{choi2020robust,mansfield1971symmetrized,burum1979analysis} still provides a convenient way to describe the pulse sequence and contributions, resulting in analytical decoupling rules for higher-order terms.
As an example, in Fig.~\ref{fig:cancellation}(c) we illustrate how first-order Magnus terms involving disorder and Heisenberg interactions have a simple geometric interpretation in analogy with dipoles, and in Fig.~\ref{fig:cancellation}(d) we illustrate how first-order Magnus terms involving Ising and Heisenberg interactions have a similar interpretation as balancing the center of mass along a given axis.
Second, we find that although there exist additional cross-terms, the majority of finite pulse duration effects can still be described as a simple extension of the effective free evolution time~\cite{choi2020robust}, making it easy to build in robustness to sequence design.
Finally, we extend the principle of pulse cycle decoupling~\cite{burum1979analysis} to more general pulse sequences and Hamiltonians, beyond those where the zeroth-order average Hamiltonian vanishes.
We use this to show how decoupling rules can be significantly simplified for pulse sequences that are composed of common motifs, such as spin echoes~\cite{hahn1950spin} or WAHUHA blocks~\cite{waugh1968approach} (Fig.~\ref{fig:cancellation}(a,b)), resulting in \textit{time-local} decoupling conditions even for higher-order Magnus contributions.
Together, these techniques allow us to find higher-order robust pulse sequences with substantially improved performance for a variety of dynamical decoupling, Hamiltonian engineering and quantum sensing applications, as discussed here and in the accompanying paper, Ref.~\cite{otherpaper}.

This paper is organized as follows: in Sec.~\ref{sec:formalism}, we review our representation of the pulse sequence and associated interaction picture Hamiltonian, as well as existing decoupling rules for the zeroth order effective Hamiltonian.
In Sec.~\ref{sec:SystematicHigherOrder}, we utilize this representation to provide general expressions for the higher-order Magnus contributions and present systematic decoupling conditions for higher-order terms.
We then analyze the structures present in these decoupling rules in Sec.~\ref{sec:HigherOrderRules}, finding significant simplifications for commonly-found pulse sequence structures in both the disorder-dominant and interaction-dominant regimes.
We also tabulate the resulting decoupling rules, and provide a pictorial depiction of them.
In Sec.~\ref{sec:SequenceSearch}, we provide further details on the efficient numerical screening of pulse sequences, resulting in high-performance pulse sequences for dynamical decoupling, quantum sensing, and quantum simulation. Finally, in Sec.~\ref{sec:conclusion} we conclude with a discussion of further extensions and future directions of the formalism.
A summary of the notation adopted in this manuscript can be found in Appendix.~\ref{sec:Conventions}.

\section{General Framework and Review of Existing Results}
\label{sec:formalism}

\subsection{General Framework and Frame Representation}

We begin by introducing our method to represent the pulse sequence and associated average Hamiltonian, which will greatly simplify the analysis of effective Hamiltonians and finite pulse effects compared to the conventional representation of individual pulses. We will adopt the toggling-frame sequence representation (also known as the Mansfield representation) used in Choi et al.~\cite{choi2020robust,mansfield1971symmetrized}, which focuses on how operators are transformed under the pulses, rather than the applied pulses themselves. In addition to being a complete and concise representation of the pulse sequence, this representation also has the additional advantage that it leads to simple decoupling conditions that are amenable to fast numerical screening.

Consider a pulse sequence composed of $n$ global spin-rotation pulses $\{P_1, \cdots, P_n\}$ acting on a system with native Hamiltonian $H$, with a free evolution time $\tau_k$ preceding the $k$th pulse $P_k$. The interaction picture Hamiltonian with respect to the ideal control pulses can then be written as
\begin{align}
    \tilde{H}(t)=U_c^\dagger(t)^{\otimes m} H U_c(t)^{\otimes m},
\end{align}
where $\tilde{H}(t)$ is the interaction picture Hamiltonian at time $t$, $U_c(t)$ is the single spin rotation due to the control field (e.g. $U_c(t)=P_kP_{k-1}\cdots P_1$ right after the $k$th pulse), and $m$ is the number of spins in the system.

Assuming ideal, infinitely fast rotation pulses, and that the combined rotation unitary is identity $P_n\cdots P_1=I$, we can write the total unitary evolution as
\begin{align}
    U(T)=\mathcal{T}\exp(-\int_0^T i\tilde{H}(t)dt)\approx \exp(-iH^{(0)}T),
\end{align}
where $\mathcal{T}$ indicates time-ordering and $T$ is the Floquet period. For pulse separations much shorter than the dynamical timescale of the system, we can conveniently write the effective Hamiltonian to leading order as
\begin{align}
\label{eq:AHT}
    H^{(0)}=\frac{1}{T}\int_0^T \tilde{H}(t_1)dt_1.
\end{align}

For general system Hamiltonians satisfying the secular approximation (rotating wave approximation under a strong quantizing field)~\cite{choi2020robust}, the interaction picture Hamiltonian $\tilde{H}_s(t)$ can be uniquely determined by transformations of the $S^z$ operator (we will refer to these as toggling ``frames")
\begin{align}
\tilde{S}^z(t)=U_c^\dagger(t)S^zU_c(t)=\sum_\mu F_{\mu}(t)S^\mu,
\end{align}
where $S^\mu$ is a basis for the spin system, e.g. the Pauli spin operators for qubits, and we have defined the coefficients
\begin{align}
    F_\mu(t)=2\textrm{Tr}[S^\mu\tilde{S}^z(t)].
\end{align}

Assuming ideal, instantaneous pulses (the case of finite pulse effects and other associated imperfections are discussed in Sec.~\ref{sec:finitepulse}), we can express the preceding information in the form of a single $4\times N$ matrix, where each element $F_{\mu,k}$ corresponds to the coefficient $F_\mu(t)$ during the $k$th free evolution time, and the last row contains the free evolution time duration.

As a concrete example, let us consider a spin-1/2 system, where each pulse $P_k$ is assumed to be a $\pi/2$ pulse around $\pm \hat{x},\hat{y}$.
Note that $\pi$ pulses can be viewed as two consecutive $\pi/2$ pulses, with zero time separation in between.
With $S^\mu$ chosen to be the Pauli basis, a spin echo can be represented as
\begin{align}
    \begin{pmatrix}
{\bf F}\\
\boldsymbol{\tau}
\end{pmatrix}_\text{echo}
=\begin{pmatrix}
0 & 0 \\
0 & 0 \\
+1 & -1 \\
\tau& \tau\\
\end{pmatrix},\label{eq:echo}
\end{align}
while the WAHUHA decoupling sequence for dipolar interactions~\cite{waugh1968approach} can be expressed as
\begin{align}
    \begin{pmatrix}
{\bf F}\\
\boldsymbol{\tau}
\end{pmatrix}_\text{WAHUHA}
=\begin{pmatrix}
0 & 0 & +1 & +1 & 0 & 0\\
0 & +1 & 0 & 0 & +1 & 0 \\
+1 & 0 & 0 & 0 & 0 & +1 \\
\tau & \tau & \tau & \tau & \tau & \tau\\
\end{pmatrix}.\label{eq:WAHUHA}
\end{align}
Pictorially, we can represent the first three rows of the matrix by the blocks in Fig.~\ref{fig:cancellation}(a,b), in which a yellow(green) block indicates a +1(-1) value along the given axis (row) at a given time (column).
We illustrate more advanced versions of spin echoes and WAHUHA blocks in Fig.~\ref{fig:cancellation}(a,b), in both pulse notation and the frame matrix notation utilized here.
In the preceding examples, we have neglected finite pulse duration effects, but they can be easily treated by specifying an additional intermediate toggling frame with zero time duration.

This representation allows us to easily express the interaction picture Hamiltonian $\tilde{H}(t)$. For example, for a spin-1/2 dipolar-interacting many-body spin system with on-site disorder, the system Hamiltonian can be written as
\begin{align}
    H_{dip}&=\sum_i h_iS_i^z+\sum_{ij}J_{ij}(S_i^xS_j^x+S_i^yS_j^y-2S_i^zS_j^z)\nonumber\\
    &=\sum_i h_iS_i^z+\sum_{ij}J_{ij}(\vec{S}_i\cdot\vec{S}_j-3S_i^zS_j^z),\label{eq:Hdip}
\end{align}
where $h_i$ is the on-site disorder strength for spin $i$, and $J_{ij}$ is the dipolar interaction between spins $i$ and $j$.
With our representation, for sequences composed of $\pi/2$ or $\pi$ pulses around $\pm\hat{x},\hat{y}$, the interaction picture Hamiltonian during the $k$th free evolution time can be easily expressed as
\begin{align}
    \tilde{H}_{dip,k}&=\sum_{i\mu} F_{\mu,k} h_i S_i^\mu+\sum_{ij}J_{ij}\vec{S}_i\cdot\vec{S}_j-3\sum_{ij\mu}F_{\mu,k}^2 J_{ij}S_i^\mu S_j^\mu,
\end{align}
where we have organized the terms according to how they transform with $F_{\mu,k}$. Using these expressions, it is easy to verify that the spin echo cancels disorder, since $\sum_k F_{\mu,k}\tau_k=0$, while the WAHUHA pulse sequence fully symmetrizes (decouples) dipolar interactions, since $\sum_{k} F_{\mu,k}^2\tau_k$ is the same for all $\mu$.

Motivated by these considerations, for any secular Hamiltonian, we will organize the interaction picture Hamiltonian in terms of how the operators transform as the toggling frame changes.
Let us write
\begin{align}
    \tilde{H}(t)=\sum_\alpha c_\alpha(t)\mathcal{O}^\alpha,
    \label{eq:OperatorExpansion}
\end{align}
where $c_\alpha(t)$ are time-dependent coefficients encoding the frame transformations of a general operator basis set $\mathcal{O}^\alpha$ (we will use greek letters to denote labels of operator sets in the remainder of the paper). For the example above in Eq.~(\ref{eq:Hdip}), we can write out the individual terms in the summation as
\begin{align}
    \mathcal{O}^0&=\sum_{ij}J_{ij}\vec{S}_i\cdot\vec{S}_j,& c_0(t)&=1,\label{eq:Operator0}\\
    \mathcal{O}^{1,\mu}&=\sum_{i}  h_i S_i^\mu,& c_{1,\mu}(t)&=F_{\mu,k},\label{eq:Operator1}\\
    \mathcal{O}^{2,\mu}&=3\sum_{ij} J_{ij}S_i^\mu S_j^\mu,& c_{2,\mu}(t)&=F_{\mu,k}^2,\label{eq:Operator2}
\end{align}
which clearly illustrates how the various terms in the Hamiltonian transform differently with the toggling frames.

\subsection{Magnus Expansion}
With this general representation framework in hand, we will now briefly review the Magnus expansion, which provides a useful tool to calculate the effective dynamics of the periodically-driven system, and extend the analysis beyond the average Hamiltonian described in Eq.~(\ref{eq:AHT}).

The total unitary over a single Floquet cycle can be expressed in terms of a time-independent effective Hamiltonian $\mathcal{U}(T)=\text{exp}(-iH_{\textit{eff}}T)$, where in the fast-driving limit, the effective Hamiltonian can be written via the Magnus expansion up to order $l$ as $H_{\textit{eff}}\approx\sum_{k=0}^lH^{(k)}$, with
\begin{align}
    H^{(0)}&=\frac1{T}\int_0^TH(t_1)dt_1,\\
    H^{(1)}&=\frac{-i}{2T}\int_0^Tdt_1\int_0^{t_1}dt_2[H(t_1),H(t_2)],\label{eq:firstorderH}\\
    H^{(2)}&=\frac1{6T}\int_0^Tdt_1\int_0^{t_1}dt_2\int_0^{t_2}dt_3\nonumber\\&\hspace{-0.3in} \left([H(t_1),[H(t_2),H(t_3)]]+[H(t_3),[H(t_2),H(t_1)]]\right).
\end{align}

Higher-order terms are more complex, involving progressively deeper nested commutators, but in the fast-driving limit they will be relatively suppressed, and we can focus on the leading order terms above.

Plugging in Eq.~(\ref{eq:OperatorExpansion}) and separating the time-independent operator commutation relation information from the time integrals, we have
\begin{align}
   H^{(0)}
   &=\sum_\alpha\frac{1}{T}\Op^\alpha \int_0^T dt_1c_\alpha (t_1),\label{eq:ZerothOrderMagnus}
    \\ H^{(1)}%&=\frac{-i}{2T}\int_0^tdt_1\int_0^{t_1}dt_2[c^\mu (t_1)\Op_\mu ,c^\nu (t_2)\Op_\nu ]\\
    &=\sum_{\alpha,\beta}\frac{-i}{2T}[\Op^\alpha ,\Op^\beta ]\int_0^Tdt_1\int_0^{t_1}dt_2c_\alpha (t_1)c_\beta (t_2),\label{eq:FirstOrderMagnus}
    \\ H^{(2)}&=\sum_{\alpha,\beta,\gamma} \frac{1}{6T}[\Op^\alpha,[\Op^\beta,\Op^\gamma]]\iiint\displaylimits_{0\le t_3\le t_2\le t_1\le T}dt_1dt_2dt_3\nonumber\\
    &\qty(c_\alpha(t_1)c_\beta(t_2)c_\gamma(t_3)+c_\alpha(t_3)c_\beta(t_2)c_\gamma(t_1))\label{eq:SecondOrderMagnus}.
\end{align}
This allows us to reduce the computation of the Magnus expansion to the evaluation of a few integrals on the $c(t)$ coefficients, which can in turn be readily phrased as algebraic conditions on the set of frame transformations.

\subsection{Review of Zeroth Order Rules}
Using the preceding framework, we can readily write down conditions for the cancellation or symmetrization of various zeroth order average Hamiltonian terms, see also Ref.~\cite{choi2020robust} for details.
For example, plugging Eqs.~(\ref{eq:Operator0}-\ref{eq:Operator2}) into Eq.~(\ref{eq:ZerothOrderMagnus}), and assuming ideal, instantaneous pulses, we can easily see that on-site disorder is cancelled when $\sum_k F_{\mu,k}\tau_k=0$ for each axis $\mu=\hat{x},\hat{y},\hat{z}$, while interactions are symmetrized into a Heisenberg Hamiltonian or cancelled when $\sum_k F_{\mu,k}^2\tau_k$ is equal for all different $\mu$.

More importantly, as shown in Ref.~\cite{choi2020robust}, these conditions can be readily generalized to the case with pulse imperfections.
The primary effect of finite pulse durations is to extend the effective free evolution times in each frame, as most of the terms generated by $\pi/2$ rotations can be written as an average of the Hamiltonian before and after the pulse.
However, there will be additional terms arising from rotation angle errors or interaction cross-terms during rotations, which give rise to additional chirality or parity conditions between neighboring frames~\cite{choi2020robust}.

The simple, time-local nature (all rules only involve neighboring frames) of these decoupling conditions enabled efficient design and screening of pulse sequences.
Indeed, using these simple decoupling conditions, novel pulse sequences with improved decoupling performance have been found, leading to the demonstration of the first solid-state AC magnetometer that operates beyond the limits of spin-spin interactions~\cite{zhou2020quantum}.
However, the further extension of such techniques to incorporate higher-order Magnus contributions and improve performance is at first sight challenging, given the time-non-local nature of higher-order Magnus terms, which involve commutators between all times of a Floquet cycle.
In the following, we demonstrate how this challenge can be systematically overcome by utilizing the structure of commonly-used pulse sequences.

\section{Systematic Analysis of Higher-Order Magnus Terms}
\label{sec:SystematicHigherOrder}

With the basic formalism in hand, we now turn to the systematic extension of these decoupling rules from zeroth-order to higher-order.
First, we will describe the pulse cycle decoupling principle~\cite{burum1979analysis} and extend it to the case of more general interactions, which serves as a useful tool to decompose \textit{non-local} higher-order terms into \textit{local} blocks.
We will then systematically derive expressions for first- and second-order Magnus contributions in the general case, assuming ideal pulses.
Finally, we briefly describe how the treatment can be readily generalized to the case with finite pulse durations, primarily by extending the effective duration of free evolution times, with more details given in Appendix.~\ref{supp:full1storder}.

\subsection{Pulse Cycle Decoupling}
\label{sec:pulsecycledecoupling}
In order to simplify sequence analysis, it is helpful to be able to break down larger pulse sequences into smaller blocks and analyze them independently.
In this section, we show how common motifs used in sequence design---in the form of spin echoes or interaction symmetrization---allow us to decompose higher-order contributions into a sum of independent, local pieces, no longer requiring non-local correlators between arbitrary locations and thus significantly simplifying the design.
Our results are applicable even to some cases where the symmetrization results in a residual Heisenberg interaction Hamiltonian, thus extending the existing methods~\cite{burum1979analysis} of pulse cycle decoupling to new and experimentally important regimes.

Let us proceed by examining the first-order contribution when sequentially applying two sequences $A$ and $B$ of equal length $T$.
We can split the first-order Magnus contribution in Eq.~(\ref{eq:firstorderH}) into integrals within the first and second sequence respectively, and cross terms between the two sequences, resulting in
\begin{align}
H^{(1)}=\frac{1}{2}H^{(1)}_A+\qty(\frac{-i}{4T})\qty[TH^{(0)}_B,TH^{(0)}_A]+\frac{1}{2}H^{(1)}_B,
\end{align}
where $H^{(0,1)}_{A,B}$ are the zeroth (first) order effective Hamiltonians during pulse sequences $A$ and $B$.

The key observation of pulse cycle decoupling is that if the commutator $\qty[H^{(0)}_A, H^{(0)}_B]$ vanishes, then the first-order contribution fully decouples into the sum of that in each individual block, regardless of the details.
In prior work~\cite{burum1979analysis}, this was achieved by making one of the average Hamiltonians vanish, thus causing the commutator to automatically vanish as well.
For more general Hamiltonians, however, this no longer directly applies, since the Heisenberg interaction is invariant under global rotations and cannot be cancelled with a global drive~\cite{choi2017dynamical}.

\begin{figure}
\begin{center}
\includegraphics[width=\columnwidth]{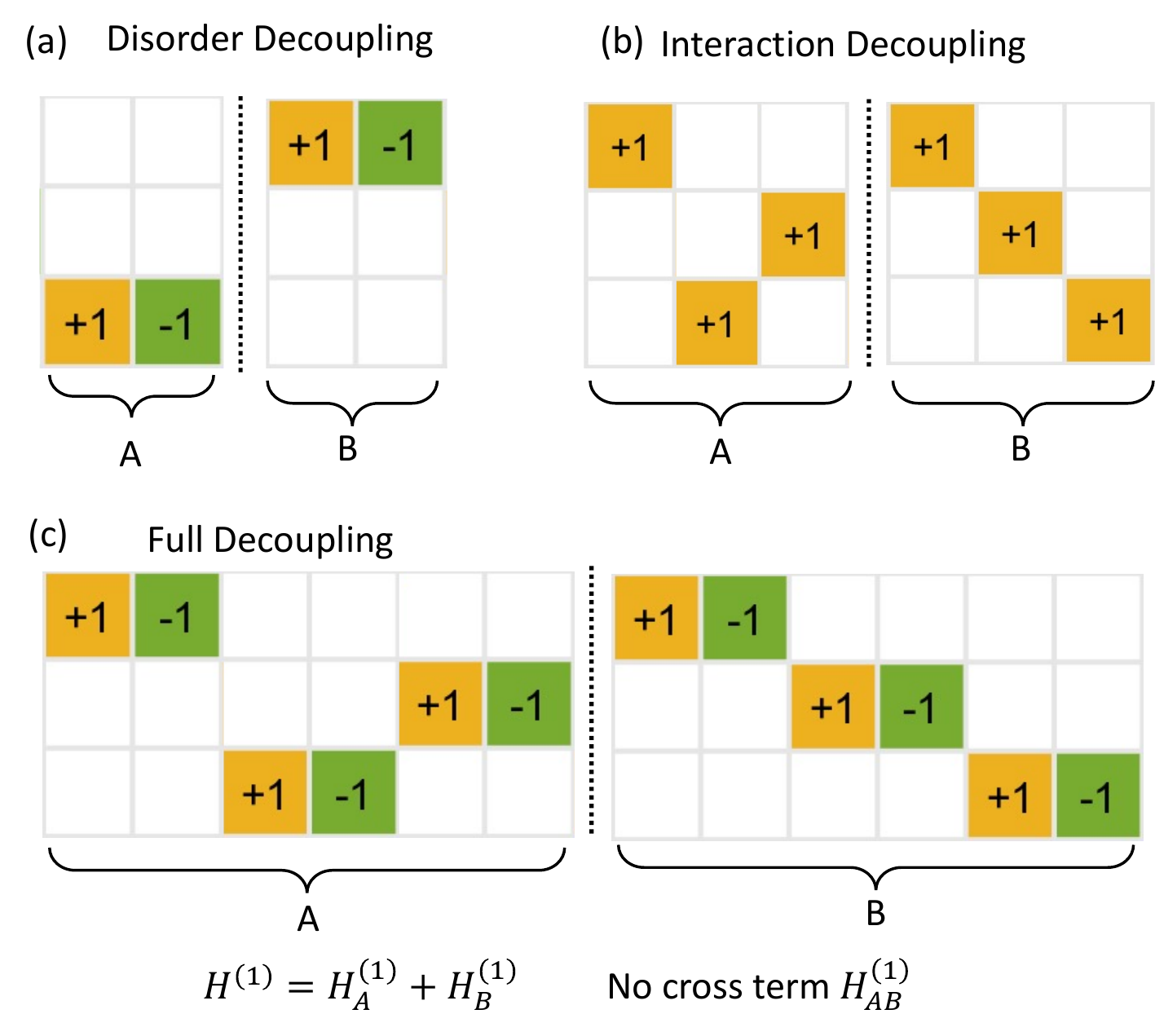}
\caption{{\bf Pulse Cycle Decoupling.} Example pulse blocks where the pulse cycle decoupling principle can be applied, and first-order contributions separate into independent, local terms.
(a) A spin echo cancels disorder, and thus $H^{(0)}_A=H^{(0)}_B=0$, satisfying the pulse cycle decoupling condition. Note that in this case, a single block being fully decoupled is sufficient for the standard pulse cycle decoupling condition to hold.
(b) Frame symmetrization between the different axes symmetrizes the interaction into a Heisenberg form, so even if the two blocks have a residual zeroth order interaction Hamiltonian, the zeroth order contributions commute, and thus the pulse cycle decoupling condition still applies.
(c) Combining spin echoes and frame symmetrization yields pulse cycle decoupling for the full Hamiltonian.}
\label{fig:PulseCycleDecoupling}
\end{center}
\end{figure}

Despite this challenge, we find that we can still make use of the pulse cycle decoupling principle in many scenarios beyond the case where the Hamiltonian vanishes.
First, even if the total Hamiltonian does not vanish, pulse cycle decoupling can still apply to individual terms.
For example, if a given block fully decouples disorder (the rapid echo blocks in Fig.~\ref{fig:cancellation}(a)), then any first-order terms involving disorder will not have cross terms between this block and other parts of the sequence, simplifying the design.
Second, if the interaction is transformed into the same form in two separate blocks, then although $H_{A,B}^{(0)}$ are both nonzero, they still commute, and so the pulse cycle decoupling principle still applies (see e.g. Fig.~\ref{fig:PulseCycleDecoupling}).
Thus, even if the interaction has a Heisenberg component that cannot be cancelled, the cross-term is still zero because $[\sum_{ij}J_{ij}S_i\cdot S_j, \sum_{ij}J_{ij}S_i\cdot S_j]=0$.
This insight generalizes the pulse cycle decoupling principle to cases in which one desires to engineer a non-zero target Hamiltonian, significantly expanding its applicability.

While we have illustrated the pulse cycle decoupling principle at first order, the same methods also apply at higher-order by generalizing the arguments in Ref.~\cite{burum1979analysis}.
For example, the second-order Magnus contribution can be expressed as a sum of commutators between zeroth and first-order terms, and thus if lower orders are fully symmetrized, then the second-order Magnus contribution will also separate into independent, local blocks.

\subsection{First-Order Decoupling}
\label{sec:firstorder}
We now return to analyze the structure of higher-order Magnus terms directly and derive decoupling rules for various contributions.
The expressions here will be derived in full generality, without making use of the pulse cycle decoupling principle, although we will use this to further simplify the expressions in following sections.

In order to better understand the structure of the first-order Magnus contributions and derive decoupling rules, let us rewrite the preceding expressions into a form that relates them to zeroth-order Magnus contributions and makes clear how terms can be cancelled.

Denoting the zeroth order contribution up to a given time as $\cc_\alpha(t)\equiv\int_0^tc_\alpha (t_1)dt_1$, we can rewrite Eq.~(\ref{eq:FirstOrderMagnus}) as
\begin{align}
    H^{(1)}&=\frac{-i}T\sum_{\alpha>\beta}[\Op^\alpha ,\Op^\beta ]\nonumber\\
    &\times\int_0^Tc_\alpha (t_1)\cc_\beta(t_1)dt_1-\frac12\cc_\alpha(T)\cc_\beta(T).
    \label{eq:firstordermagnusinitial}
\end{align}

Focusing first on the case of instantaneous, ideal pulses (the more general case will be treated in Sec.~\ref{sec:finitepulse}), our toggling-frame Hamiltonian becomes piecewise-constant in time, allowing us to replace integrals with summations.
We can then define the discrete frame equivalents of the terms in the previous section, letting $c_{\alpha,k}=c_\alpha (t)$ for $t\in[t_k-\tau_k/2,t_k+\tau_k/2]$ and $\cc_{\alpha,k}\equiv\sum_{l<k}c_{\alpha,l}\tau_l$, and find
\begin{align}
H^{(1)}&=\sum_{\beta<\alpha}[\Op^\alpha ,\Op^\beta ]\nonumber\\&\times\qty(\sum_{k=1}^n c_{\alpha,k}\tau_k\qty(\cc_{\beta, k}+\frac12\tau_k c_{\beta,k})-\frac12\cc_{\alpha,n+1}\cc_{\beta,n+1}),
\label{eq:firstordermagnusfinal}
\end{align}
The first term in the parenthesis can be interpreted as a product between the $c_\alpha$ coefficient during a given frame and the integral of zeroth order average Hamiltonians up to the center of the frame.
The last term is simply the product of two zeroth-order contributions over the entire Floquet period, which will vanish when zeroth-order decoupling rules are satisfied.
Geometrically, this expression can be understood as rewriting the triangular integration area in Eq.~(\ref{eq:FirstOrderMagnus}) into a sum over thin column slices.

We emphasize that these results apply to all first-order contributions, illustrating the common structure found in the decoupling of many different types of terms.
By keeping track of the running sum, the evaluation of this expression now requires only linear time, as opposed to the naive quadratic complexity.
In addition, although Eq.~(\ref{eq:FirstOrderMagnus}) still contains products of coefficients that are non-local in time, in Sec.~\ref{sec:HigherOrderRules} we shall see that in many cases of interest, it can be reduced into simple, local decoupling rules.

We also generalize these results to the case with finite pulse durations in Sec.~\ref{sec:finitepulse}.
The primary effect, similar to the zeroth-order case~\cite{choi2020robust}, is to lengthen the effective duration of each free evolution time by an amount proportional to the pulse duration.
There will be additional cross terms that we tabulate in the appendix, but they are generally smaller.

\subsection{Second-Order Decoupling}
\label{sec:secondorder}
We can now apply the same formalism to the second-order Magnus contributions.
As we show in Appendix.~\ref{supp:secondorder}, by reordering the integrals, we can re-express the second-order contribution in terms of the zeroth- and first-order contributions at different times. Let us define the first-order contribution from time $t_1$ to $t_2$ of the operator $[\Op^\beta,\Op^\gamma]$ as
\begin{align}
c^{(1)}_{\beta, \gamma}(t_1,t_2)
&=\substack{\iint\\t_1<t_b<t_a<t_2}\qty(c_{\beta}(t_a)c_{\gamma}(t_b)-c_{\beta}(t_b)c_{\gamma}(t_a)),
\end{align}
where we have dropped the integrand $dt_a dt_b$ for notational simplicity here and below.
We can rewrite the expression as follows
\begin{align}
    H^{(2)}&=\frac1{6T}\sum_{\alpha<\beta<\gamma} ([\Op^\alpha,[\Op^\beta,\Op^\gamma]])\nonumber\\
    \times&\int_0^Tdt_1c_\alpha(t_1)\left(c^{(1)}_{\beta,\gamma}(0,t_1)-c^{(1)}_{\beta,\gamma}(t_1,T) \right).
\end{align}

Thus, we see a very similar structure as at first order, wherein the second-order term can also be expressed as a simple integral of lower-order products, enabling formulation of simple decoupling rules.

\subsection{Robustness Conditions}
\label{sec:finitepulse}
We now extend these results to the case with finite pulse durations, and describe how to incorporate robustness to these effects into the sequence design.
We will focus our attention on the dominant contribution, which we find to be a simple extension of the effective free evolution time by an amount proportional to the pulse duration.
A full treatment of the finite pulse effects, including additional sub-leading cross terms, can be found in Appendix.~\ref{supp:full1storder}.

\begin{figure}
\begin{center}
\includegraphics[width=\columnwidth]{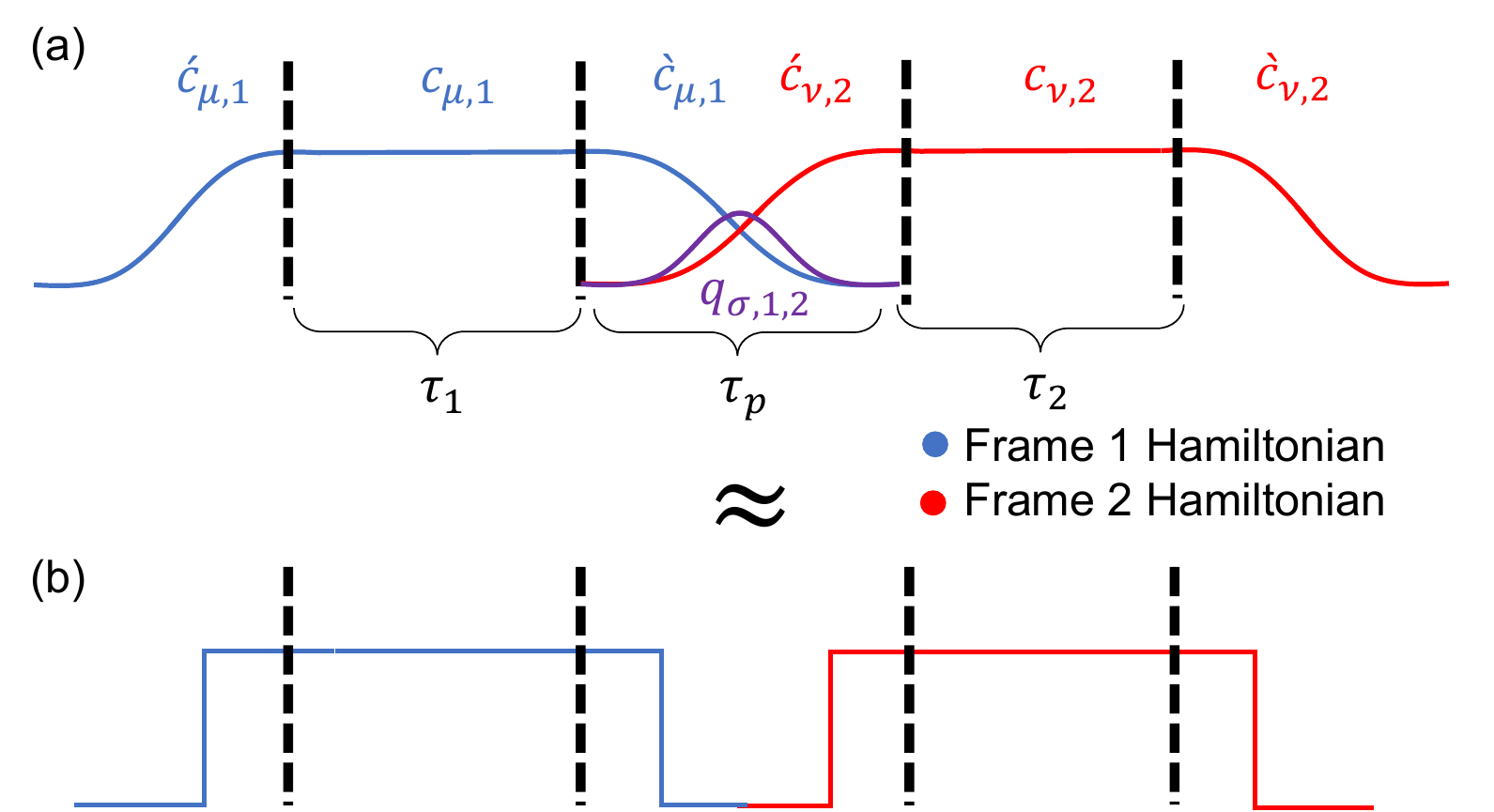}
\caption{{\bf Evolution of frames with finite pulse durations.} (a) Illustration of the coefficients of two different frames in a typical sequence.
With finite pulse durations, the coefficients have additional tails on the two sides as well as cross-terms in overlapping regions.
(b) The leading order effect of this is an extension of the free evolution durations, while still treating each frame independently.}
\label{fig:PulseTerm}
\end{center}
\end{figure}

As shown in Fig.~\ref{fig:PulseTerm}, with a finite pulse duration, the coefficient $c_\alpha(t)$ of each term of the Hamiltonian will consist of a ramp up (the preceding rotation), free evolution in the frame, a ramp down (the following rotation), as well as some additional cross-terms between the frames.

The primary effect of finite pulse effects is illustrated in Fig.~\ref{fig:PulseTerm}(b), in which the effective duration of each frame is lengthened by the integral of $\cpl_{\alpha, k}$ and $\cm_{\alpha , k}$ over time.
This is a simple extension of the calculation for the zeroth-order case, and scales as $O(\tau \tau_p)$, where $\tau$ is the free evolution time and $\tau_p$ is the pulse duration.
We incorporate this into the main term in the decoupling rule table, Tab.~\ref{tab:FirstOrder}, as described in more detail in the next section.

In addition to this dominant term, there are contributions that scale as $O(\tau_p^2)$ or higher powers.
First, we have to treat the overlap of the frames, which gives rise to the additional term:
\begin{align}
\iint\displaylimits_{0<\theta_2<\theta_1<\pi/2}\cm_{\alpha,k}(\theta_1)\cpl_{\beta,k+1}(\theta_2)-\cpl_{\alpha,k+1}(\theta_2)\cm_{\beta,k}(\theta_1).
\label{eq:finitepulse}
\end{align}
These terms correspond to the overlap of the ramp up of one frame with the ramp down of another. We note that we get a positive effect from the ramp down of one into the ramp up of the other, and a negative effect from the ramp up of one with the ramp down of the other. These are the main finite pulse effects to each of the correction terms, and are shown in column 4 of Tab.~\ref{tab:FirstOrder}. 

The final contribution originates from first-order contributions involving interaction cross-terms $q_\rho$.
More specifically, during the continuous rotation from an $XX$ Hamiltonian to a $YY$ Hamiltonian, $XY$-type terms are generated; the $q_\rho$ cross-terms come from first-order cross-terms between these $XY$-type terms and other terms.
Note that there will be no such cross-terms for the disorder part of our Hamiltonian.
Thus, although the magnitude of this term can in principle scale as $O(\tau_p\tau)$, in practice the coefficients are small for disorder-dominated systems, and we will analyze this in detail instead in Appendix~\ref{supp:full1storder}.

\section{Higher-Order Decoupling Rules}
\label{sec:HigherOrderRules}
\subsection{Summary of General Rules}

\setlength\extrarowheight{5pt}

\begin{table*}
\centering
{\footnotesize
\begin{tabular}{|c|c|c|c|c|}
\hline\# & Decoupling Effect & Algebraic Condition (Ideal Pulse) & Algebraic Condition (Finite Pulse Correction) & Local Cancellation Condition \\[5pt]\hline
1 & Disorder-Disorder & $2 \sum_{k=1}^n F_{\mu, k}\left(\tau_k+\frac{4}{\pi} \tau_p\right) F_{<k}^v-\bar{F}^\mu \bar{F}^v$ & $\left(\frac{2 \tau_p}{\pi}\right)^2\left(1-\frac{\pi}{4}\right) \sum_{k=1}^\eta F_{\mu, k} F_{v, k+1}-F_{v, k} F_{\mu, k+1}$ & Fast Echo\\[5pt]\hline
2 & Disorder-Ising & 2 $\sum_{k=1}^n F_{\mu, k}\left(\tau_k+\frac{4}{\pi} \tau_p\right) I_{<k}^v-\bar{F}^\mu \bar{I}^v$ & $\left(\frac{2 \pi_p}{\pi}\right)^2\left(\frac{\pi}{4}-\frac{2}{3}\right) \sum_{k=1}^n F_{\mu, k}\left|F_{v, k+1}\right|-\left|F_{v, k}\right| F_{\mu, k+1}$ & Fast Echo\\[5pt]\hline
3 & Ising-Ising & $\sum_{k=1}^n\left|F_{\mu, k}\right|\left(\tau_k+\tau_p\right)\left(I_{<k}^v-I_{>k}^v\right)$ & $\left(\frac{2 \tau_p}{\pi}\right)^2\left(\frac{\pi^2}{32}-\frac{1}{4}\right) \sum_{k=1}^n\left|F_{\mu, k}\right|\left|F_{v, k+1}\right|-\left|F_{v, k}\right|\left|F_{\mu, k+1}\right|$ & Block Symmetrization\\[5pt]\hline
4 & Disorder-Heisenberg & $\sum_{k=1}^n F_{\mu, k}\left(\tau_k+\frac{4}{\pi} \tau_p\right)\left(t_k-\frac{T}{2}\right)$ & 0 & Dipole Cancellation\\[5pt]\hline
5 & Ising-Heisenberg & $\sum_{k=1}^n |F_{\mu, k}|\left(\tau_k+\frac{4}{\pi} \tau_p\right)\left(t_k-\frac{T}{2}\right)$ & 0 & Row Balancing\\[5pt]\hline
6 & Heisenberg-Heisenberg & 0 & 0 & Automatically Satisfied\\[5pt]\hline
7 & 2nd-Order Disorder & $2 \sum_{k=1}^n F_{\mu, k}\left(\tau_k+\frac{1}{\pi} \tau_p\right) F_{<k}^{\nu, \rho}-F^\mu F^{\nu, \rho}$ & & Fast Echo\\[5pt]\hline
    
\end{tabular}
}
\caption{{\bf Higher-order cancellation rules.}
    Each row describes a different higher-order contribution, described in the second column.
    The third column gives the expressions that need to vanish to cancel the largest contributions to a given term, including contributions from free evolution periods and frame lengthening corrections due to finite pulse durations.
    The fourth column gives further finite pulse corrections.
    The $q$-cross terms are ignored, only giving an additional correction term in terms involving the Ising interaction.
    The last column gives the local cancellation condition that eliminates that type of error. $F_{<k}^\nu=\sum_{l<k} F_{\nu,l} (\tau_l+\frac4\pi\tau_p)$, $I_{<k}^\nu=\sum_{l<k} |F_{\nu,l}| (\tau_l+\tau_p)$, $\bar{F}^\nu=\sum_k F_{\nu,k} (\tau_k+\frac4\pi\tau_p)$, $\bar{I}^\nu=\sum_k |F_{\nu,k}| (\tau_k+\tau_p)$.}
    \label{tab:FirstOrder}
\end{table*}

\begin{comment}
\begin{table*}
\begin{center}
    \includegraphics[width=2\columnwidth]{figures/Main table 2.png}
    \caption{{\bf Higher-order cancellation rules.}
    %
    Each row describes a different higher-order contribution, described in the second column.
    %
    The third column gives the expressions that need to vanish to cancel the largest contributions to a given term, including contributions from free evolution periods and frame lengthening corrections due to finite pulse durations.
    %
    The fourth column gives further finite pulse corrections.
    %
    The $q$-cross terms are ignored, only giving an additional correction term in terms involving the Ising interaction.
    %
    The last column gives the local cancellation condition that eliminates that type of error. $F_{<k}^\nu=\sum_{l<k} F_{\nu,l} (\tau_l+\frac4\pi\tau_p)$, $I_{<k}^\nu=\sum_{l<k} |F_{\nu,l}| (\tau_l+\tau_p)$, $\bar{F}^\nu=\sum_k F_{\nu,k} (\tau_k+\frac4\pi\tau_p)$, $\bar{I}^\nu=\sum_k |F_{\nu,k}| (\tau_k+\tau_p)$.}
    \label{tab:FirstOrder}
\end{center}
\end{table*}
\end{comment}

We now utilize the results from the preceding section to derive concrete decoupling rules for various important higher-order contributions.

Plugging in different Hamiltonian terms into the expressions derived in Sec.~\ref{sec:firstorder},\ref{sec:secondorder}, we arrive at the decoupling rules in Tab.~\ref{tab:FirstOrder}.
As higher-order terms originate from commutators between different terms, we label the cancellation rules with all operators involved.
The table includes two types of contributions: first, there are the main terms that will appear even with ideal, infinitesimally short pulses, together with corrections to their effective duration due to finite pulse durations; second, we include terms that come purely from the finite pulse duration, in the form of the overlapping pulse term derived in Eq.~(\ref{eq:finitepulse}).
There is one additional type of term, as mentioned in the preceding section, that involves interaction cross terms $q_\rho$ during continuous rotations.
We omit them from this table, since they do not appear for disorder terms that are dominant in our experiments, but we discuss them in more detail in Appendix~\ref{supp:full1storder}.
Note also that since the Magnus expansion is not invariant with respect to cyclic permutations of the pulse sequence, there are modifications to terms relating to the first and last frames in the complete expression.
However, we neglect them from this table, both to simplify notation, as well as due to the fact that after many Floquet cycles we expect the contributions from these boundary terms to be diminished.

While the decoupling rules are somewhat more complicated than the zeroth-order rules derived in Ref.~\cite{choi2020robust}, many of them nonetheless have simple geometric intuitions (Fig.~\ref{fig:cancellation}), and can be further simplified for many common scenarios.
Moreover, the decoupling rules can often be satisfied with simple local motifs, further simplifying the pulse sequence design task.
For example, in many cases, by using the pulse cycle decoupling principle described in Sec.~\ref{sec:pulsecycledecoupling}, one can apply the same rules as zeroth-order sequence design, except requiring the cancellation on a much faster timescale.
These considerations are summarized in the last column of Tab.~\ref{tab:FirstOrder}.
We will now go through a few representative examples in more detail, and explain how to interpret and simplify the rules.

\subsection{Fast Echo Cancellation for Disorder-Related Terms}
Let us now focus our attention on first-order rules related to disorder-disorder and disorder-Ising terms.

To derive the first and second conditions in Tab.~\ref{tab:FirstOrder}, which hold with full generality and no restrictions on the frame matrix, let us examine the structure of the first-order Magnus contribution shown in Eq.~(\ref{eq:firstordermagnusfinal}).
There, we found that a generic first-order Magnus contribution can be rewritten as a product between the current frame contribution of one term and the cumulative contribution of another term, together with a factor corresponding to the total zeroth order contribution of both terms.
Plugging the disorder and Ising expressions in Eqs.~(\ref{eq:Operator0}-\ref{eq:Operator2}) into Eq.~(\ref{eq:firstordermagnusfinal}) results in the rules in the third column of Tab.~\ref{tab:FirstOrder}.

Due to this common structure, we see in Tab.~\ref{tab:FirstOrder} that the decoupling of first-order disorder-disorder and disorder-Ising contributions are almost identical, except replacing one term from scaling with $F_{\mu,k}$ to be scaling instead as $F_{\mu,k}^2=|F_{\mu,k}|$ for $F_{\mu,k}=0,\pm 1$.
We can thus decouple the primary contribution of both of these first-order effects with the same pulse sequence block, simply by arranging the frames to form fast spin echoes.
These fast spin echoes are illustrated in Fig.~\ref{fig:cancellation}(a).
To see this, first note that the commutator pre-factor implies that there will be a nonzero contribution only when $\mu\neq \nu$.
With a spin echo block, the contribution from $F_{\mu,k}$ flips and cancels, while the sum $F_{<k}^\nu$ or $I_{<k}^\nu$ is along a different axis and thus remains unchanged during this time.
Moreover, the average zeroth order disorder Hamiltonian over the entire pulse sequence will also vanish due to the spin echo blocks.
Thus, the main term rules in conditions 1 and 2 are satisfied in Tab.~\ref{tab:FirstOrder}.

In column 4 of Tab.~\ref{tab:FirstOrder}, we derived additional corrections to the expressions, originating from the pulse-induced overlaps in Eq.~(\ref{eq:finitepulse}).
Interestingly, we find again that the different terms share some common structures, where they can be related to each other simply by replacing $F_{\nu,k}$ by $|F_{\nu,k}|$.
Moreover, we find that the finite pulse duration corrections for first-order disorder-disorder terms are proportional to that of rotation angle errors at zeroth-order~\cite{choi2020robust}, making it automatically satisfied if the latter has been incorporated into sequence design.

We also include an example of a second-order rule involving disorder only in Tab.~\ref{tab:FirstOrder}.
As one can see, the structure bears many similarities with the first-order contributions.
In the case where the pulse sequence is composed of fast echoes, we can further simplify the expressions.

\subsection{Block Symmetrization for Ising-Ising Terms}
Moving on to the first-order Ising-Ising terms (row 3 of Tab.~\ref{tab:FirstOrder}), we see that the structure of the expression again has many similarities as above.
However, here we have grouped the terms slightly differently, since the zeroth-order sum $\bar{I}^\nu=\sum_k |F_k^\nu|(\tau_k+\tau_p)$ will always be nonzero.
Expressed in this way, first-order Ising-Ising interactions are decoupled by ensuring that for every pair of axes $\mu$ and $\nu$, the cumulative occurrences of $\nu$ frames before and after each $\mu$ frame are equal, i.e. the appearance of the two frames is balanced.

Based on this result, we find that a simple motif is to perform a mirror symmetrization~\cite{mansfield1971symmetrized} of the frames within each block, as illustrated in Fig.~\ref{fig:cancellation}(b).
Note that since the coefficient of the Ising contribution is identical regardless of the sign of the frame, only the relative frame ordering matters and not the sign.
If within each block the frames are balanced along the $\hat{x}$, $\hat{y}$ and $\hat{z}$ directions, and mirror symmetrization in terms of ordering is performed, then using the pulse cycle decoupling principle, the first-order Ising-Ising contribution will be cancelled.
Here, contrary to global mirror symmetrization, we find that symmetrization within local blocks can also be a useful tool to effectively cancel certain first-order contributions.

The finite pulse duration correction terms for the first-order Ising-Ising contribution are shown in the fourth column of the Tab.~\ref{tab:FirstOrder}.
They resemble the other correction terms, but with additional absolute value signs, and can be easily incorporated as decoupling rules in a similar fashion.
 
\subsection{Dipole Cancellation and Row Balancing for Heisenberg-Related Terms}
Let us now examine terms related to the Heisenberg interaction $\vec{S}\cdot \vec{S}$.
As noted in Eq.~(\ref{eq:Operator0}), the Heisenberg interaction is invariant under frame transformations.
Thus, the first-order expression resulting from the Heisenberg term and a different term will be the commutator between a constant term (Heisenberg), and a term that depends on the frame transformations (other).
The inclusion of the time integrals then result in the expressions in row 4-6 of Tab.~\ref{tab:FirstOrder}, where there are no first-order contributions between two Heisenberg Hamiltonians because they are identical and thus commute.
In this case, because the Heisenberg Hamiltonian is invariant, the frame length extension becomes exact, and we do not need to include any additional finite pulse corrections in the table.

To explore in more detail what the resulting rules mean, let us first recall the expressions for cancelling disorder at zeroth order.
Here, we found that in order for disorder to be cancelled at zeroth order, we require the average disorder along each axis to vanish, i.e.
\begin{align}
    \sum_{k=1}^n F_{\mu,k}(\tau_k+\frac{4}{\pi}\tau_p)=0.
\end{align}
A useful physical analogy to interpret this expression is to associate a positive(negative) charge with $F_{\mu,a}=+1$($-1$).
The zeroth-order decoupling condition then dictates that the average charge is 0.

Generalizing the analogy to first-order terms, the first-order term of a given Hamiltonian contribution with the invariant Heisenberg Hamiltonian will be proportional to the given Hamiltonian, weighted by its location in time in the sequence.
This is because of the integration limits in Eq.~(\ref{eq:FirstOrderMagnus}), where the relative ordering of the time variable values of the two Hamiltonians determines the sign of the expression.
Furthering the electromagnetic analogy, this results in a distance weighting factor from the center of the pulse sequence timing.
Thus, the first-order expression resembles the expression of a dipole, with charge given by $F_{\mu,k}$ at each time point and distance being the distance in time to the center of the sequence.
Cancelling this contribution requires the net dipole along each axis to vanish, as illustrated in Fig.~\ref{fig:cancellation}(c).
We note that this intuition was key to improving decoupling pulse sequence performance in Sec.~\ref{sec:finalsequence}, and led to insights regarding the dichotomy between AC field sensing and decoupling for existing pulse sequences described in Ref.~\cite{otherpaper}.

Similarly, we can also analyze the expression for cross-terms between Ising interactions and Heisenberg interactions, simply by replacing the general charge by a non-negative charge value ($|F_{\mu,k}|$).
With only positive charges, the condition can also be alternatively viewed geometrically as balancing frame weights in each row;
as illustrated in Fig.~\ref{fig:cancellation}(d), one can imagine a fulcrum placed at the middle of a sequence, and for a given axis, placing a weight whenever the frame is along this axis (regardless of it being a positive or negative frame); the rule then becomes that the row would balance.

\section{Detailed Sequence Design Procedure}
\label{sec:SequenceSearch}

We now utilize the preceding insights to design higher-order pulse sequences for various applications, focusing on the case of interacting spin ensembles dominated by on-site disorder~\cite{kucsko2018critical,zhou2020quantum}.
The result is a pulse sequence that decouples all zeroth-order and first-order contributions in the Magnus expansion, and is robust against disorder to second order, which we name DROID-R2D2 (Disorder RObust Interaction Decoupling - Robust To Disorder 2nd order), and a pulse sequence that achieves similar results but also has interesting AC field sensing capabilities~\cite{otherpaper}.
These pulse sequences were crucial for a variety of our recent experiments in dynamical decoupling, quantum metrology~\cite{otherpaper}, and Hamiltonian engineering~\cite{martin2022controlling}.
We will illustrate the complete design procedure in detail, and mention a few practical tricks to improve the efficiency of sequence screening and to examine larger design spaces of pulse sequences.

\begin{enumerate}
    \item [1.] \textbf{Choose target decoupling rules}
\end{enumerate}

The first step is to determine the set of decoupling rules that should be satisfied by the desired sequence.
The choice of this set is usually informed by several factors: First, the target application may influence which terms need to be decoupled.
For example, if we wish to study many-body dynamics in a disordered system, we may wish to preserve the disorder term while engineering interactions.
Alternatively, if we are interested in quantum sensing, then there may be additional design rules that are imposed to maximize sensitivity.

\begin{figure}
    \centering
    \includegraphics[width=\columnwidth]{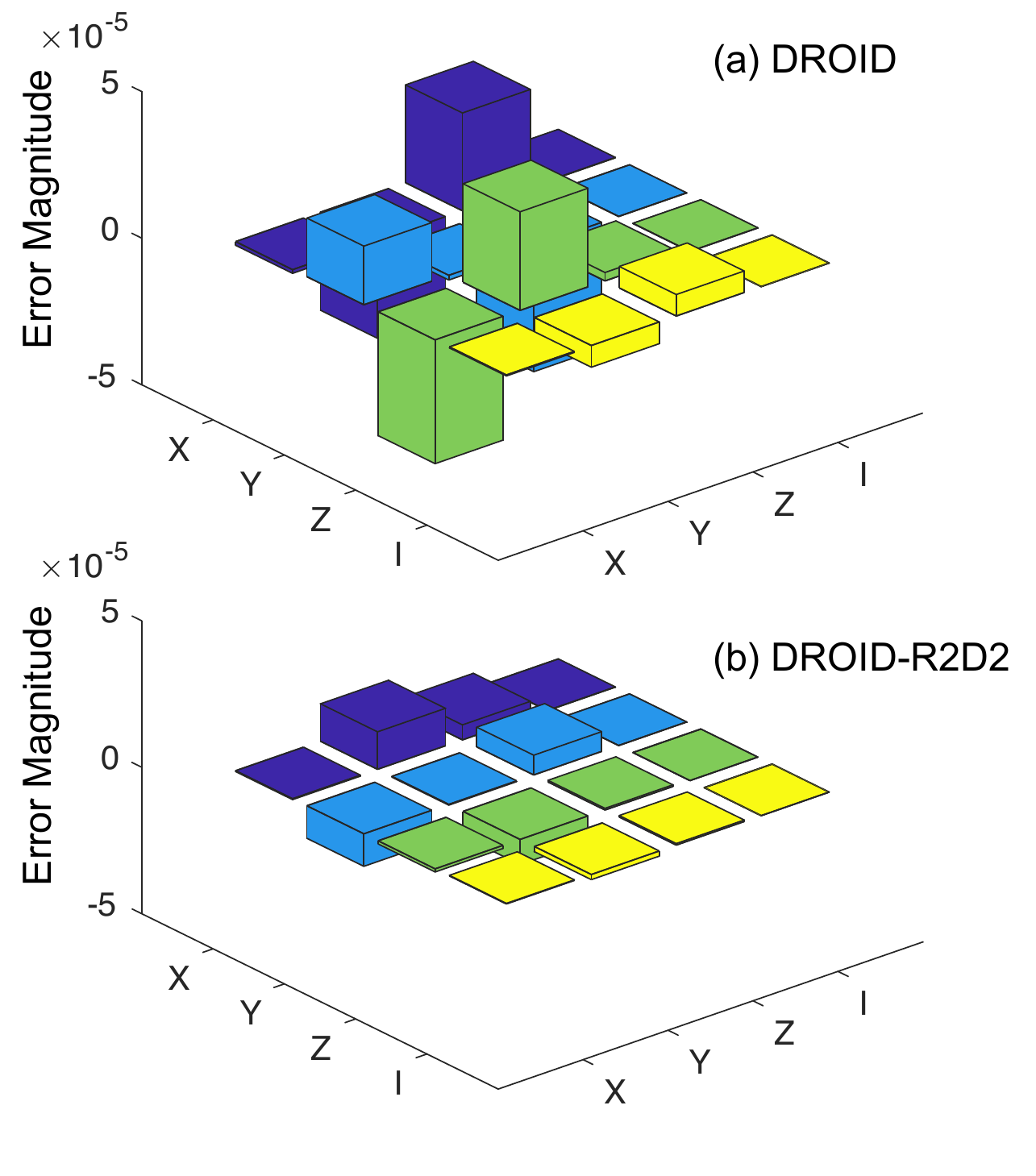}
    \caption{{\bf Comparison between decoupling sequence performance.} We compare the effective error Hamiltonian extracted from the matrix log of a two-spin unitary for disorder strengths $2\pi\times(4,0)$ MHz and interactions $2\pi\times 35$ kHz, for the pulse sequences DROID (a) and DROID-R2D2 (b).
    The ideal Heisenberg portion of the Hamiltonian has been subtracted out.
    Bars represent coefficients of the effective Hamiltonian decomposed in the Pauli basis.
    We find that the error is much smaller for our new sequence DROID-R2D2.}
    \label{fig:seq_comparison}
\end{figure}

\begin{figure}
    \centering
    \includegraphics[width=\columnwidth]{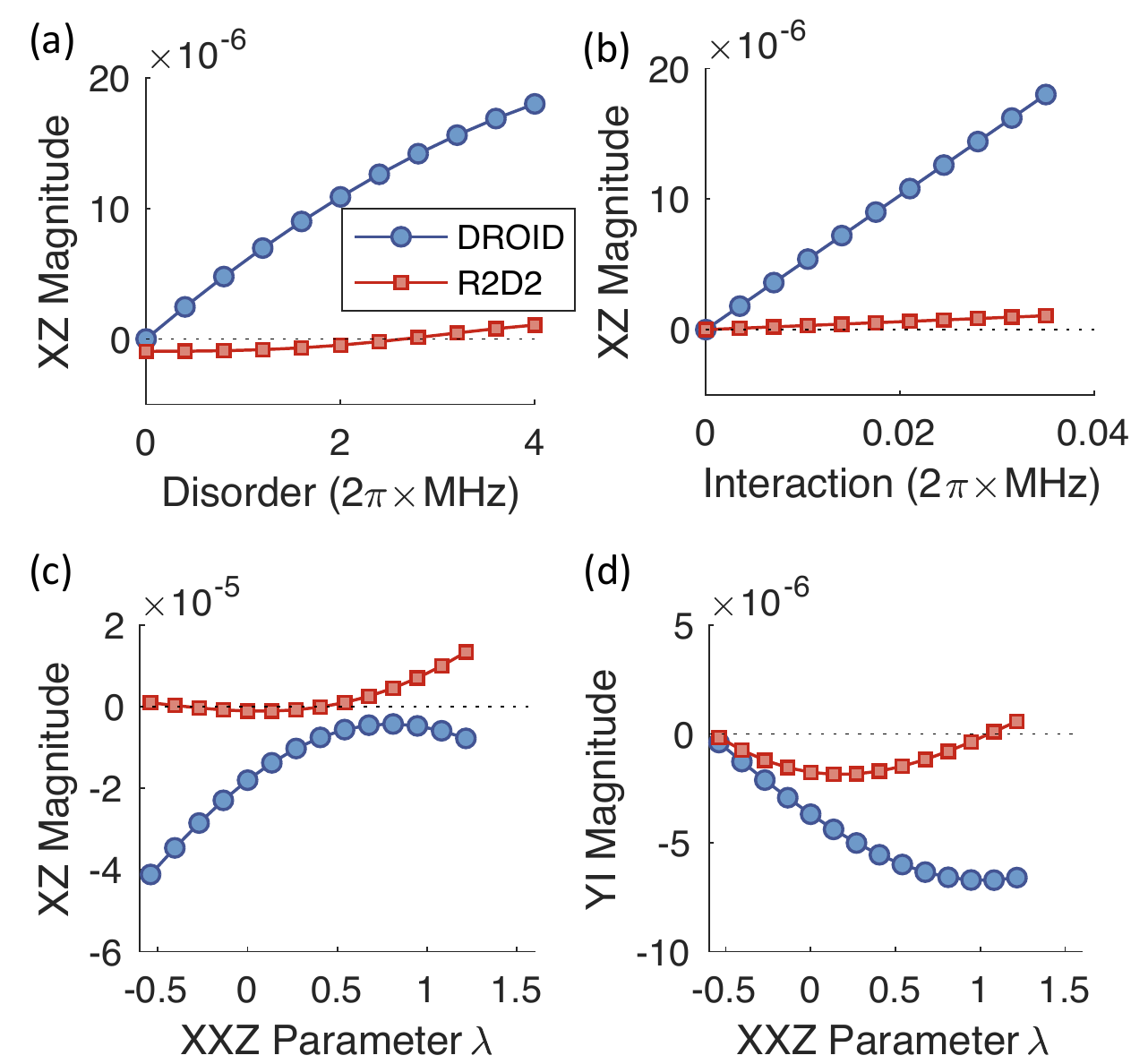}
    \caption{{\bf Error scaling for various pulse sequences.} (a,b) Magnitude of XZ error term in the effective Hamiltonian as a function of disorder (a) and interactions (b) for the DROID sequence (blue circles) and the DROID-R2D2 sequence (red squares, much smaller errors). The dominant scaling is linear for DROID. (c,d) Magnitude of XZ (c) and YI (d) error terms in the effective Hamiltonian for different XXZ Hamiltonians, where DROID again shows a systematically larger error and stronger Hamiltonian dependence.}
    \label{fig:error_scaling}
\end{figure}

Second, the experimental system characteristics may inform which contributions are most important to decouple.
As an example, dense electronic spin ensembles, such as nitrogen-vacancy (NV) centers and nitrogen (P1) defects in diamond, or rare earth ions, typically have much larger disorder than interactions~\cite{zhou2020quantum,zu2021emergent,merkel2021dynamical}.
Thus, it is much more important to address disorder-related effects to higher-order than interaction-related effects.
The relative importance of different contributions can be made more quantitative by using our expressions for various terms to estimate the typical total magnitude of each of the Hamiltonian terms.

Finally, for a given pulse sequence, we can also diagnose the dominant residual term by examining a cluster of a few spins, typically two or three, and computing the exact unitary for a set of disorder and interaction values.
Taking the matrix log of the unitary yields the exact effective Hamiltonian, and performing polynomial fits of the dominant terms with respect to the disorder and interaction strengths informs us which type of contribution is the largest, as well as the order at which it contributes in the Magnus expansion.
For example, in Fig.~\ref{fig:seq_comparison}(a), we find that the dominant error terms for the existing DROID sequence from Ref.~\cite{zhou2020quantum} are the XZ and ZX components of the Hamiltonians, when decomposed in the Pauli basis.
In Fig.~\ref{fig:error_scaling}(a,b), we find that the dominant scaling of this term is linear in both the disorder strength and interaction strength, suggesting that it originates from a first-order cross-term between them.
This motivated us to systematically include decoupling rules that target this effect.

In practice, we search for sequences by randomly enumerating those of a fixed length that satisfy a chosen set of rules (see below for a description of how to efficiently enforce rules).
We iterate the preceding error diagnosis step several times by identifying the dominant contributions for typical pulse sequences, and adding in new rules to fully decouple them.
Each addition of a new dominant rule eliminates the most poorly performing sequences, and increases the probability of enumerating a sequence with high coherence time; see Fig.~2 of the accompanying paper~\cite{otherpaper}.

Following this procedure leads us to include the following decoupling rules for our disorder-dominated NV center ensemble: decoupling of all zeroth-order conditions, as described in Ref.~\cite{choi2020robust}; decoupling of all first-order conditions involving at least one factor of disorder, including disorder-disorder cross-terms, disorder-Ising and disorder-Heisenberg cross-terms, for both free evolution times and pulses; second-order disorder-disorder-disorder cross-terms, for both free evolution times and pulses.

\begin{enumerate}
    \item [2.] \textbf{Efficiently construct candidate frame sets}
\end{enumerate}

With the set of target decoupling rules in hand, we now discuss how to efficiently enumerate pulse sequences satisfying a set of imposed decoupling rules.
The number of possible frame sets without any additional constraints is combinatorially large.
For example, even a sequence consisting of 12 free evolution times connected by $\pi/2$ pulses, including intermediate frames for the finite pulse durations (e.g. the frame half way through $\pi$ pulses or composite $\pi/2$ pulses), admits approximately $4^{23}\approx 10^{14}$ distinct pulse sequences (each frame is connected to 4 other frames by $\pi/2$ pulses, and the first frame is fixed to be $+\hat{z}$).
However the vast majority of these sequences will not satisfy our rules. 
Therefore it is essential to enumerate only sequences that satisfy them.

For the disorder-dominated interacting NV ensembles we work with, we choose to impose the following structures to efficiently pre-screen pulse sequences:
we require that all frames, including both free evolution and pulse frames, come in spin echo pairs, in order to echo out disorder on the fastest possible timescale.
In addition, we require an equal number of elements along each row, so as to symmetrize interactions.
Finally, we impose the ``dipole" rules for first-order disorder-Heisenberg cross-terms, by requiring there to be an equal number of $+-$ and $-+$ spin echoes.

In order to directly restrict the search space to candidate frame sequences that satisfy the above rules, we separately enumerate the locations of X, Y and Z spin echo pairs, and enumerate the echo ordering signs (i.e. whether the echo frames has the ordering $+-$ or $-+$) of both free evolution frames and finite pulse frames.
We then combine these pieces of information to generate candidate frame sequences, imposing the additional constraint that each frame must be distinct from the two neighboring frames, to ensure that a $\pi/2$ pulse is applied and the pulse error calculation is accurate.

\begin{enumerate}
    \item [3.] \textbf{Screen frame sets using decoupling rules}
\end{enumerate}

Having generated candidate frame sets that already have a number of rules enforced by construction, we now proceed to screen through them by applying the remaining decoupling rules.
In order to speed up the screening process, the key insight is to transform the original rules into a vectorized form, such that fast matrix computation can be performed, significantly reducing the run time.
This is achieved by labeling the frames as 1 to 6 for $+x$,$+y$,$+z$,$-x$,$-y$,$-z$, and noting that the rules become simple cumulative sums of index matching results when expressed in this fashion.
We then further simplify them based on known decoupling structures (e.g. the rapid spin echoes built into the sequence).
Moreover, when evaluating some of the higher-order expressions in full generality, we can keep track of the cumulative integral of lower-order terms~\cite{khaneja2005optimal}, which reduces the time complexity of computing many such terms to linear in the sequence length, rather than a higher polynomial scaling.

\begin{enumerate}
    \item [4.] \textbf{Verify performance and further optimization}
\end{enumerate}

To optimize the performance of the pulse sequences, we further symmetrize the pulse sequence to reduce higher-order error contributions.
Here, for dynamical decoupling, we employ the symmetrization used in Ref.~\cite{choi2020robust}, where the frames are repeated twice, but the frame ordering is reversed and sign of all frames flipped in the second repetition.
For the quantum metrology pulse sequences designed here and in Ref.~\cite{otherpaper}, this symmetrization will affect the magnetic field sensitivity, and consequently we employ a mirror-symmetrization instead, where the frame ordering is reversed but the sign is not flipped in the second repetition.

Finally, we numerically simulate the performance of these pulse sequences to identify the ones with the longest decoupling timescales.
Effective Hamiltonian extraction using the matrix log of the unitary can identify dominant error terms for the pulse sequences employed, and the whole design procedure can be repeated with an improved rule set.
For the above final set of decoupling rules, we no longer find a single contribution that dominates over the others, instead seeing a competition between several different contributions.

\begin{figure*}
    \centering
    \includegraphics[width=2\columnwidth]{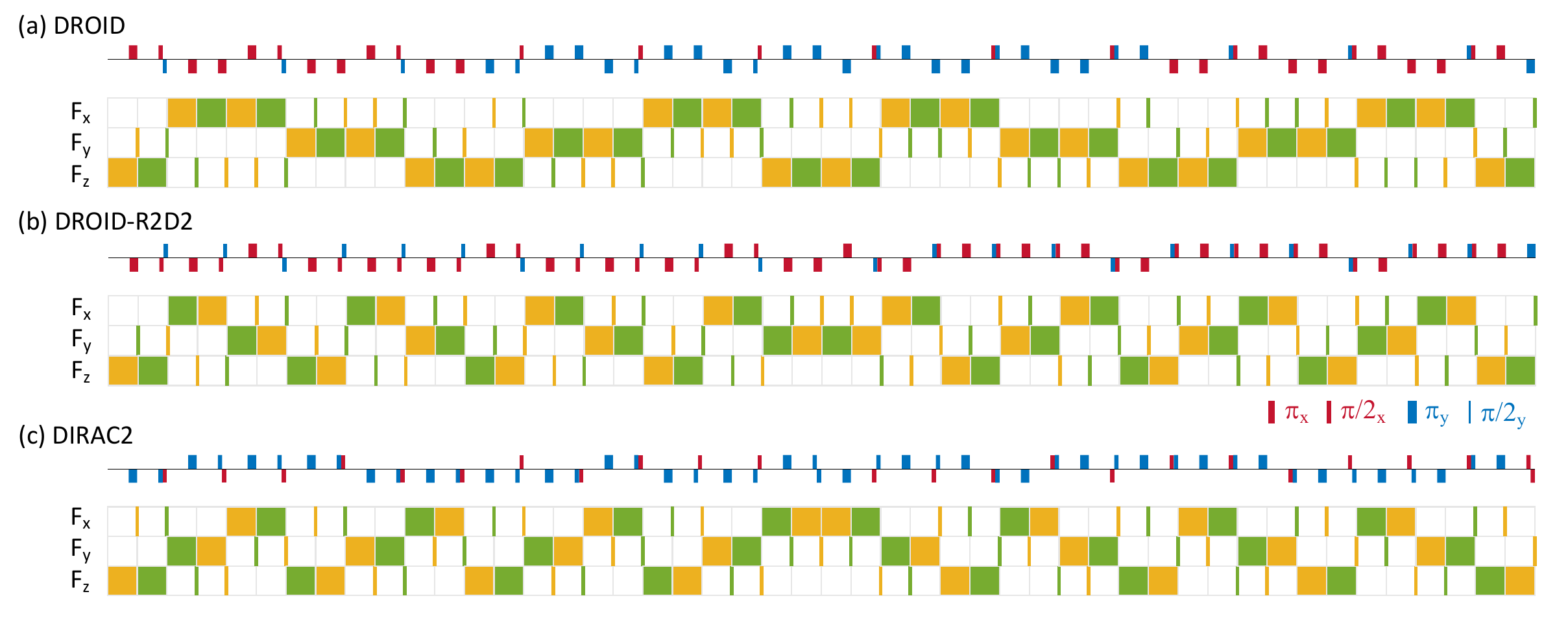}
    \caption{{\bf Details of higher-order pulse sequences.} Pulse representation (top) and frame representation (bottom) of various pulse sequences found using the higher-order design rules developed in this paper.
    (a) DROID~\cite{zhou2020quantum}, the previous best pulse sequence for dynamical decoupling and quantum metrology in disorder-dominated interacting spin ensembles.
    (b) DROID-R2D2, a new pulse sequence designed for dynamical decoupling and Hamiltonian engineering, in which all first-order contributions are fully cancelled and disorder is cancelled to second order.
    See Ref.~\cite{martin2022controlling} for an application of this pulse sequence to many-body XXZ dynamics.
    (c) DIRAC2, a new pulse sequence that has similar characteristics, but is designed for improved quantum sensing, see Ref.~\cite{otherpaper} for details and experimental demonstration of the improved sensitivity compared to the best known sequences.}
    \label{fig:full_sequence}
\end{figure*}

\subsection{Resulting Pulse Sequences}
\label{sec:finalsequence}
Using the decoupling rules described above, we designed pulse sequences for dynamical interaction decoupling, many-body physics, and quantum metrology.

For dynamical decoupling and Hamiltonian engineering, one of the best pulse sequences we identified, DROID-R2D2, is shown in Fig.~\ref{fig:full_sequence}(b).
We find that compared to the previous best pulse sequence DROID~\cite{zhou2020quantum}, as shown in Fig.~\ref{fig:full_sequence}(a), that had significant residual first-order cross terms (Fig.~\ref{fig:seq_comparison}(a)), primarily cross terms between disorder and first-order Heisenberg interactions, the residual errors when examining the effective Hamiltonian are much reduced (Fig.~\ref{fig:seq_comparison}(b)).

Moreover, we can adapt this pulse sequence to perform Hamiltonian engineering by adjusting the frame durations along the $\hat{x}$, $\hat{y}$ and $\hat{z}$ axes~\cite{martin2022controlling}, resulting in a tunable interaction Hamiltonian
\begin{align}
H_{XXZ}=\sum_{ij}J_{ij}\qty[\frac{1+\lambda}{3}\qty(S_i^xS_j^x+S_i^yS_j^y)+\frac{1-2\lambda}{3}S_i^zS_j^z],
\end{align}
where $\lambda$ is a coefficient that tunes the XXZ Hamiltonian.
We find that our techniques also significantly reduce the error in engineering a wide range of generic XXZ Hamiltonians.
In Fig.~\ref{fig:error_scaling}(c,d), we see that both two-body and single-body imperfection terms are much smaller across a wide range of different Hamiltonians, which can help improve the fidelity of Hamiltonian engineering and reduce systematic artifacts.
These techniques can be readily generalized to engineer XYZ Hamiltonians, with different coefficients in front of each term, or even more complex many-body Hamiltonians.

The same techniques can also be used to design pulse sequences for improved quantum sensing, as we explain in more detail in Ref.~\cite{otherpaper}.
The key insight is that current pulse sequences for quantum sensing~\cite{zhou2020quantum}, which periodically flip the spin along each axis with the same frequency as the target signal, will always result in a violation of the ``net dipole cancellation" rule in Fig.~\ref{fig:cancellation}(c) for first-order disorder-Heisenberg terms.
This imposes a fundamental trade-off between sensitivity and decoupling quality for current pulse sequences.
With this insight from higher-order decoupling rules, we are able to design the new pulse sequence DIRAC2 (DIsorder Robust AC sensing with period 2), as shown in Fig.~\ref{fig:full_sequence}(c), which circumvents this issue by targeting a sensing signal half the frequency of frame flipping, thereby fully cancelling all first-order Magnus contributions while also increasing the rate of spin echo decoupling, leading to better performance.
See Ref.~\cite{otherpaper} for a more detailed description.

\section{Discussion and Conclusion}
\label{sec:conclusion}
We have developed a general framework for dynamical Hamiltonian engineering that includes higher-order considerations.
Contrary to the naive expectation, we found that many higher-order decoupling conditions can still have simple, intuitive interpretations, particularly when the pulse sequence is designed to have certain structures in it.
We analytically derived a number of decoupling rules for higher-order contributions, and used them to design robust pulse sequences in disorder-dominated systems for dynamical decoupling, Hamiltonian engineering, and quantum sensing, significantly improving upon state-of-the-art pulse sequences.

While we have focused on the application of our techniques to the case of electronic spin ensembles, where disorder is much larger than spin-spin interactions, we believe that our techniques can be applied to disparate systems such as NMR~\cite{peng2021deep,cory1990time}, simply by changing which rules are emphasized and included at higher-order.
It may also be interesting to further extend the techniques to even higher-order than the ones that we have considered here~\cite{hohwy1997elimination}, or to examine alternative expansions beyond the Magnus expansion~\cite{mote2016five,mananga2016on}.
In our formalism, the contributions from higher-order terms are decomposed into an operator commutation portion, and a portion that relates to the frame matrix and pre-factors.
This also makes the extension to higher-spin systems relatively straightforward, and can be combined with recent methods for robust Hamiltonian engineering with higher-spin systems~\cite{zhou2023robust,leitao2022qudit}.
With these further improvements, we believe that our framework presents a key tool for advanced Hamiltonian engineering pulse sequence design, with broad applications in dynamical decoupling, quantum many-body physics, and quantum metrology.

\section*{Acknowledgements}
We thank J.~Choi, A.~Douglas, H.~Gao, N.~Maskara, P.~Peng, M.~Yu for helpful discussions. This work was supported in part by CUA, HQI, NSSEFF, ARO MURI, DARPA DRINQS, Moore Foundation GBMF-4306, NSF PHY-1506284.

\bibliography{main}

\begin{thebibliography}{42}%
\makeatletter
\providecommand \@ifxundefined [1]{%
 \@ifx{#1\undefined}
}%
\providecommand \@ifnum [1]{%
 \ifnum #1\expandafter \@firstoftwo
 \else \expandafter \@secondoftwo
 \fi
}%
\providecommand \@ifx [1]{%
 \ifx #1\expandafter \@firstoftwo
 \else \expandafter \@secondoftwo
 \fi
}%
\providecommand \natexlab [1]{#1}%
\providecommand \enquote  [1]{``#1''}%
\providecommand \bibnamefont  [1]{#1}%
\providecommand \bibfnamefont [1]{#1}%
\providecommand \citenamefont [1]{#1}%
\providecommand \href@noop [0]{\@secondoftwo}%
\providecommand \href [0]{\begingroup \@sanitize@url \@href}%
\providecommand \@href[1]{\@@startlink{#1}\@@href}%
\providecommand \@@href[1]{\endgroup#1\@@endlink}%
\providecommand \@sanitize@url [0]{\catcode `\\12\catcode `\$12\catcode
  `\&12\catcode `\#12\catcode `\^12\catcode `\_12\catcode `\%12\relax}%
\providecommand \@@startlink[1]{}%
\providecommand \@@endlink[0]{}%
\providecommand \url  [0]{\begingroup\@sanitize@url \@url }%
\providecommand \@url [1]{\endgroup\@href {#1}{\urlprefix }}%
\providecommand \urlprefix  [0]{URL }%
\providecommand \Eprint [0]{\href }%
\providecommand \doibase [0]{http://dx.doi.org/}%
\providecommand \selectlanguage [0]{\@gobble}%
\providecommand \bibinfo  [0]{\@secondoftwo}%
\providecommand \bibfield  [0]{\@secondoftwo}%
\providecommand \translation [1]{[#1]}%
\providecommand \BibitemOpen [0]{}%
\providecommand \bibitemStop [0]{}%
\providecommand \bibitemNoStop [0]{.\EOS\space}%
\providecommand \EOS [0]{\spacefactor3000\relax}%
\providecommand \BibitemShut  [1]{\csname bibitem#1\endcsname}%
\let\auto@bib@innerbib\@empty
%</preamble>
\bibitem [{\citenamefont {Ladd}\ \emph {et~al.}(2010)\citenamefont {Ladd},
  \citenamefont {Jelezko}, \citenamefont {Laflamme}, \citenamefont {Nakamura},
  \citenamefont {Monroe},\ and\ \citenamefont {O'Brien}}]{ladd2010quantum}%
  \BibitemOpen
  \bibfield  {author} {\bibinfo {author} {\bibfnamefont {T.~D.}\ \bibnamefont
  {Ladd}}, \bibinfo {author} {\bibfnamefont {F.}~\bibnamefont {Jelezko}},
  \bibinfo {author} {\bibfnamefont {R.}~\bibnamefont {Laflamme}}, \bibinfo
  {author} {\bibfnamefont {Y.}~\bibnamefont {Nakamura}}, \bibinfo {author}
  {\bibfnamefont {C.}~\bibnamefont {Monroe}}, \ and\ \bibinfo {author}
  {\bibfnamefont {J.~L.}\ \bibnamefont {O'Brien}},\ }\href {\doibase
  10.1038/nature08812} {\bibfield  {journal} {\bibinfo  {journal} {Nature}\
  }\textbf {\bibinfo {volume} {464}},\ \bibinfo {pages} {45} (\bibinfo {year}
  {2010})}\BibitemShut {NoStop}%
\bibitem [{\citenamefont {Vandersypen}\ and\ \citenamefont
  {Chuang}(2005)}]{vandersypen2005nmr}%
  \BibitemOpen
  \bibfield  {author} {\bibinfo {author} {\bibfnamefont {L.~M.~K.}\
  \bibnamefont {Vandersypen}}\ and\ \bibinfo {author} {\bibfnamefont {I.~L.}\
  \bibnamefont {Chuang}},\ }\href {\doibase 10.1103/RevModPhys.76.1037}
  {\bibfield  {journal} {\bibinfo  {journal} {Reviews of Modern Physics}\
  }\textbf {\bibinfo {volume} {76}},\ \bibinfo {pages} {1037} (\bibinfo {year}
  {2005})}\BibitemShut {NoStop}%
\bibitem [{\citenamefont {Degen}\ \emph {et~al.}(2017)\citenamefont {Degen},
  \citenamefont {Reinhard},\ and\ \citenamefont
  {Cappellaro}}]{degen2017quantum}%
  \BibitemOpen
  \bibfield  {author} {\bibinfo {author} {\bibfnamefont {C.~L.}\ \bibnamefont
  {Degen}}, \bibinfo {author} {\bibfnamefont {F.}~\bibnamefont {Reinhard}}, \
  and\ \bibinfo {author} {\bibfnamefont {P.}~\bibnamefont {Cappellaro}},\
  }\href {\doibase 10.1103/RevModPhys.89.035002} {\bibfield  {journal}
  {\bibinfo  {journal} {Reviews of Modern Physics}\ }\textbf {\bibinfo {volume}
  {89}},\ \bibinfo {pages} {035002} (\bibinfo {year} {2017})}\BibitemShut
  {NoStop}%
\bibitem [{\citenamefont {Georgescu}\ \emph {et~al.}(2014)\citenamefont
  {Georgescu}, \citenamefont {Ashhab},\ and\ \citenamefont
  {Nori}}]{georgescu2014quantum}%
  \BibitemOpen
  \bibfield  {author} {\bibinfo {author} {\bibfnamefont {I.~M.}\ \bibnamefont
  {Georgescu}}, \bibinfo {author} {\bibfnamefont {S.}~\bibnamefont {Ashhab}}, \
  and\ \bibinfo {author} {\bibfnamefont {F.}~\bibnamefont {Nori}},\ }\href
  {\doibase 10.1103/RevModPhys.86.153} {\bibfield  {journal} {\bibinfo
  {journal} {Reviews of Modern Physics}\ }\textbf {\bibinfo {volume} {86}},\
  \bibinfo {pages} {153} (\bibinfo {year} {2014})}\BibitemShut {NoStop}%
\bibitem [{\citenamefont {Bloch}\ \emph {et~al.}(2012)\citenamefont {Bloch},
  \citenamefont {Dalibard},\ and\ \citenamefont
  {Nascimb{\`{e}}ne}}]{bloch2012quantum}%
  \BibitemOpen
  \bibfield  {author} {\bibinfo {author} {\bibfnamefont {I.}~\bibnamefont
  {Bloch}}, \bibinfo {author} {\bibfnamefont {J.}~\bibnamefont {Dalibard}}, \
  and\ \bibinfo {author} {\bibfnamefont {S.}~\bibnamefont {Nascimb{\`{e}}ne}},\
  }\href {\doibase 10.1038/nphys2259} {\bibfield  {journal} {\bibinfo
  {journal} {Nature Physics}\ }\textbf {\bibinfo {volume} {8}},\ \bibinfo
  {pages} {267} (\bibinfo {year} {2012})}\BibitemShut {NoStop}%
\bibitem [{\citenamefont {Waugh}\ \emph {et~al.}(1968)\citenamefont {Waugh},
  \citenamefont {Huber},\ and\ \citenamefont {Haeberlen}}]{waugh1968approach}%
  \BibitemOpen
  \bibfield  {author} {\bibinfo {author} {\bibfnamefont {J.~S.}\ \bibnamefont
  {Waugh}}, \bibinfo {author} {\bibfnamefont {L.~M.}\ \bibnamefont {Huber}}, \
  and\ \bibinfo {author} {\bibfnamefont {U.}~\bibnamefont {Haeberlen}},\ }\href
  {\doibase 10.1103/PhysRevLett.20.180} {\bibfield  {journal} {\bibinfo
  {journal} {Physical Review Letters}\ }\textbf {\bibinfo {volume} {20}},\
  \bibinfo {pages} {180} (\bibinfo {year} {1968})}\BibitemShut {NoStop}%
\bibitem [{\citenamefont {Burum}\ and\ \citenamefont
  {Rhim}(1979)}]{burum1979analysis}%
  \BibitemOpen
  \bibfield  {author} {\bibinfo {author} {\bibfnamefont {D.~P.}\ \bibnamefont
  {Burum}}\ and\ \bibinfo {author} {\bibfnamefont {W.~K.}\ \bibnamefont
  {Rhim}},\ }\href {\doibase 10.1063/1.438385} {\bibfield  {journal} {\bibinfo
  {journal} {The Journal of Chemical Physics}\ }\textbf {\bibinfo {volume}
  {71}},\ \bibinfo {pages} {944} (\bibinfo {year} {1979})}\BibitemShut
  {NoStop}%
\bibitem [{\citenamefont {Cory}\ \emph
  {et~al.}(1990{\natexlab{a}})\citenamefont {Cory}, \citenamefont {Miller},\
  and\ \citenamefont {Garroway}}]{cory1990time}%
  \BibitemOpen
  \bibfield  {author} {\bibinfo {author} {\bibfnamefont {D.~G.}\ \bibnamefont
  {Cory}}, \bibinfo {author} {\bibfnamefont {J.~B.}\ \bibnamefont {Miller}}, \
  and\ \bibinfo {author} {\bibfnamefont {A.~N.}\ \bibnamefont {Garroway}},\
  }\href {\doibase 10.1016/0022-2364(90)90380-R} {\bibfield  {journal}
  {\bibinfo  {journal} {Journal of Magnetic Resonance (1969)}\ }\textbf
  {\bibinfo {volume} {90}},\ \bibinfo {pages} {205} (\bibinfo {year}
  {1990}{\natexlab{a}})}\BibitemShut {NoStop}%
\bibitem [{\citenamefont {Choi}\ \emph {et~al.}(2020)\citenamefont {Choi},
  \citenamefont {Zhou}, \citenamefont {Knowles}, \citenamefont {Landig},
  \citenamefont {Choi},\ and\ \citenamefont {Lukin}}]{choi2020robust}%
  \BibitemOpen
  \bibfield  {author} {\bibinfo {author} {\bibfnamefont {J.}~\bibnamefont
  {Choi}}, \bibinfo {author} {\bibfnamefont {H.}~\bibnamefont {Zhou}}, \bibinfo
  {author} {\bibfnamefont {H.~S.}\ \bibnamefont {Knowles}}, \bibinfo {author}
  {\bibfnamefont {R.}~\bibnamefont {Landig}}, \bibinfo {author} {\bibfnamefont
  {S.}~\bibnamefont {Choi}}, \ and\ \bibinfo {author} {\bibfnamefont {M.~D.}\
  \bibnamefont {Lukin}},\ }\href {\doibase 10.1103/PHYSREVX.10.031002}
  {\bibfield  {journal} {\bibinfo  {journal} {Physical Review X}\ }\textbf
  {\bibinfo {volume} {10}},\ \bibinfo {pages} {031002} (\bibinfo {year}
  {2020})}\BibitemShut {NoStop}%
\bibitem [{\citenamefont {Viola}\ \emph {et~al.}(1999)\citenamefont {Viola},
  \citenamefont {Knill},\ and\ \citenamefont {Lloyd}}]{viola1999dynamical}%
  \BibitemOpen
  \bibfield  {author} {\bibinfo {author} {\bibfnamefont {L.}~\bibnamefont
  {Viola}}, \bibinfo {author} {\bibfnamefont {E.}~\bibnamefont {Knill}}, \ and\
  \bibinfo {author} {\bibfnamefont {S.}~\bibnamefont {Lloyd}},\ }\href
  {\doibase 10.1103/PhysRevLett.82.2417} {\bibfield  {journal} {\bibinfo
  {journal} {Physical Review Letters}\ }\textbf {\bibinfo {volume} {82}},\
  \bibinfo {pages} {2417} (\bibinfo {year} {1999})}\BibitemShut {NoStop}%
\bibitem [{\citenamefont {Khodjasteh}\ and\ \citenamefont
  {Lidar}(2005)}]{khodjasteh2005fault}%
  \BibitemOpen
  \bibfield  {author} {\bibinfo {author} {\bibfnamefont {K.}~\bibnamefont
  {Khodjasteh}}\ and\ \bibinfo {author} {\bibfnamefont {D.~A.}\ \bibnamefont
  {Lidar}},\ }\href {\doibase 10.1103/PhysRevLett.95.180501} {\bibfield
  {journal} {\bibinfo  {journal} {Physical Review Letters}\ }\textbf {\bibinfo
  {volume} {95}},\ \bibinfo {pages} {180501} (\bibinfo {year}
  {2005})}\BibitemShut {NoStop}%
\bibitem [{\citenamefont {Uhrig}(2007)}]{uhrig2007keeping}%
  \BibitemOpen
  \bibfield  {author} {\bibinfo {author} {\bibfnamefont {G.~S.}\ \bibnamefont
  {Uhrig}},\ }\href {\doibase 10.1103/PhysRevLett.98.100504} {\bibfield
  {journal} {\bibinfo  {journal} {Physical Review Letters}\ }\textbf {\bibinfo
  {volume} {98}},\ \bibinfo {pages} {100504} (\bibinfo {year}
  {2007})}\BibitemShut {NoStop}%
\bibitem [{\citenamefont {{\'{A}}lvarez}\ \emph {et~al.}(2015)\citenamefont
  {{\'{A}}lvarez}, \citenamefont {Suter},\ and\ \citenamefont
  {Kaiser}}]{alvarez2015localization}%
  \BibitemOpen
  \bibfield  {author} {\bibinfo {author} {\bibfnamefont {G.~A.}\ \bibnamefont
  {{\'{A}}lvarez}}, \bibinfo {author} {\bibfnamefont {D.}~\bibnamefont
  {Suter}}, \ and\ \bibinfo {author} {\bibfnamefont {R.}~\bibnamefont
  {Kaiser}},\ }\href {\doibase 10.1126/science.1261160} {\bibfield  {journal}
  {\bibinfo  {journal} {Science}\ }\textbf {\bibinfo {volume} {349}},\ \bibinfo
  {pages} {846} (\bibinfo {year} {2015})}\BibitemShut {NoStop}%
\bibitem [{\citenamefont {Wei}\ \emph {et~al.}(2018)\citenamefont {Wei},
  \citenamefont {Ramanathan},\ and\ \citenamefont
  {Cappellaro}}]{wei2018exploring}%
  \BibitemOpen
  \bibfield  {author} {\bibinfo {author} {\bibfnamefont {K.~X.}\ \bibnamefont
  {Wei}}, \bibinfo {author} {\bibfnamefont {C.}~\bibnamefont {Ramanathan}}, \
  and\ \bibinfo {author} {\bibfnamefont {P.}~\bibnamefont {Cappellaro}},\
  }\href {\doibase 10.1103/PhysRevLett.120.070501} {\bibfield  {journal}
  {\bibinfo  {journal} {Physical Review Letters}\ }\textbf {\bibinfo {volume}
  {120}},\ \bibinfo {pages} {070501} (\bibinfo {year} {2018})}\BibitemShut
  {NoStop}%
\bibitem [{\citenamefont {Wei}\ \emph {et~al.}(2019)\citenamefont {Wei},
  \citenamefont {Peng}, \citenamefont {Shtanko}, \citenamefont {Marvian},
  \citenamefont {Lloyd}, \citenamefont {Ramanathan},\ and\ \citenamefont
  {Cappellaro}}]{wei2019emergent}%
  \BibitemOpen
  \bibfield  {author} {\bibinfo {author} {\bibfnamefont {K.~X.}\ \bibnamefont
  {Wei}}, \bibinfo {author} {\bibfnamefont {P.}~\bibnamefont {Peng}}, \bibinfo
  {author} {\bibfnamefont {O.}~\bibnamefont {Shtanko}}, \bibinfo {author}
  {\bibfnamefont {I.}~\bibnamefont {Marvian}}, \bibinfo {author} {\bibfnamefont
  {S.}~\bibnamefont {Lloyd}}, \bibinfo {author} {\bibfnamefont
  {C.}~\bibnamefont {Ramanathan}}, \ and\ \bibinfo {author} {\bibfnamefont
  {P.}~\bibnamefont {Cappellaro}},\ }\href {\doibase
  10.1103/PhysRevLett.123.090605} {\bibfield  {journal} {\bibinfo  {journal}
  {Physical Review Letters}\ }\textbf {\bibinfo {volume} {123}},\ \bibinfo
  {pages} {090605} (\bibinfo {year} {2019})}\BibitemShut {NoStop}%
\bibitem [{\citenamefont {Hayes}\ \emph {et~al.}(2014)\citenamefont {Hayes},
  \citenamefont {Flammia},\ and\ \citenamefont
  {Biercuk}}]{hayes2014programmable}%
  \BibitemOpen
  \bibfield  {author} {\bibinfo {author} {\bibfnamefont {D.}~\bibnamefont
  {Hayes}}, \bibinfo {author} {\bibfnamefont {S.~T.}\ \bibnamefont {Flammia}},
  \ and\ \bibinfo {author} {\bibfnamefont {M.~J.}\ \bibnamefont {Biercuk}},\
  }\href {\doibase 10.1088/1367-2630/16/8/083027} {\bibfield  {journal}
  {\bibinfo  {journal} {New Journal of Physics}\ }\textbf {\bibinfo {volume}
  {16}},\ \bibinfo {pages} {083027} (\bibinfo {year} {2014})}\BibitemShut
  {NoStop}%
\bibitem [{\citenamefont {Ajoy}\ and\ \citenamefont
  {Cappellaro}(2013)}]{ajoy2013quantum}%
  \BibitemOpen
  \bibfield  {author} {\bibinfo {author} {\bibfnamefont {A.}~\bibnamefont
  {Ajoy}}\ and\ \bibinfo {author} {\bibfnamefont {P.}~\bibnamefont
  {Cappellaro}},\ }\href {\doibase 10.1103/PhysRevLett.110.220503} {\bibfield
  {journal} {\bibinfo  {journal} {Physical Review Letters}\ }\textbf {\bibinfo
  {volume} {110}},\ \bibinfo {pages} {220503} (\bibinfo {year}
  {2013})}\BibitemShut {NoStop}%
\bibitem [{\citenamefont {Choi}\ \emph
  {et~al.}(2017{\natexlab{a}})\citenamefont {Choi}, \citenamefont {Yao},\ and\
  \citenamefont {Lukin}}]{choi2017dynamical}%
  \BibitemOpen
  \bibfield  {author} {\bibinfo {author} {\bibfnamefont {S.}~\bibnamefont
  {Choi}}, \bibinfo {author} {\bibfnamefont {N.~Y.}\ \bibnamefont {Yao}}, \
  and\ \bibinfo {author} {\bibfnamefont {M.~D.}\ \bibnamefont {Lukin}},\ }\href
  {\doibase 10.1103/PhysRevLett.119.183603} {\bibfield  {journal} {\bibinfo
  {journal} {Physical Review Letters}\ }\textbf {\bibinfo {volume} {119}},\
  \bibinfo {pages} {183603} (\bibinfo {year} {2017}{\natexlab{a}})}\BibitemShut
  {NoStop}%
\bibitem [{\citenamefont {Haas}\ \emph {et~al.}(2019)\citenamefont {Haas},
  \citenamefont {Puzzuoli}, \citenamefont {Zhang},\ and\ \citenamefont
  {Cory}}]{haas2019engineering}%
  \BibitemOpen
  \bibfield  {author} {\bibinfo {author} {\bibfnamefont {H.}~\bibnamefont
  {Haas}}, \bibinfo {author} {\bibfnamefont {D.}~\bibnamefont {Puzzuoli}},
  \bibinfo {author} {\bibfnamefont {F.}~\bibnamefont {Zhang}}, \ and\ \bibinfo
  {author} {\bibfnamefont {D.~G.}\ \bibnamefont {Cory}},\ }\href {\doibase
  10.1088/1367-2630/ab4525} {\bibfield  {journal} {\bibinfo  {journal} {New
  Journal of Physics}\ }\textbf {\bibinfo {volume} {21}} (\bibinfo {year}
  {2019}),\ 10.1088/1367-2630/ab4525}\BibitemShut {NoStop}%
\bibitem [{\citenamefont {Rose}\ \emph {et~al.}(2018)\citenamefont {Rose},
  \citenamefont {Haas}, \citenamefont {Chen}, \citenamefont {Jeon},
  \citenamefont {Lauhon}, \citenamefont {Cory},\ and\ \citenamefont
  {Budakian}}]{rose2018high}%
  \BibitemOpen
  \bibfield  {author} {\bibinfo {author} {\bibfnamefont {W.}~\bibnamefont
  {Rose}}, \bibinfo {author} {\bibfnamefont {H.}~\bibnamefont {Haas}}, \bibinfo
  {author} {\bibfnamefont {A.~Q.}\ \bibnamefont {Chen}}, \bibinfo {author}
  {\bibfnamefont {N.}~\bibnamefont {Jeon}}, \bibinfo {author} {\bibfnamefont
  {L.~J.}\ \bibnamefont {Lauhon}}, \bibinfo {author} {\bibfnamefont {D.~G.}\
  \bibnamefont {Cory}}, \ and\ \bibinfo {author} {\bibfnamefont
  {R.}~\bibnamefont {Budakian}},\ }\href {\doibase 10.1103/PhysRevX.8.011030}
  {\bibfield  {journal} {\bibinfo  {journal} {Physical Review X}\ }\textbf
  {\bibinfo {volume} {8}},\ \bibinfo {pages} {011030} (\bibinfo {year}
  {2018})}\BibitemShut {NoStop}%
\bibitem [{\citenamefont {Slichter}(2013)}]{slichter2013principles}%
  \BibitemOpen
  \bibfield  {author} {\bibinfo {author} {\bibfnamefont {C.~P.}\ \bibnamefont
  {Slichter}},\ }\href@noop {} {\emph {\bibinfo {title} {{Principles of
  magnetic resonance}}}},\ Vol.~\bibinfo {volume} {1}\ (\bibinfo  {publisher}
  {Springer Science \& Business Media},\ \bibinfo {year} {2013})\BibitemShut
  {NoStop}%
\bibitem [{\citenamefont {Mehring}(2012)}]{mehring2012principles}%
  \BibitemOpen
  \bibfield  {author} {\bibinfo {author} {\bibfnamefont {M.}~\bibnamefont
  {Mehring}},\ }\href@noop {} {\emph {\bibinfo {title} {{Principles of high
  resolution NMR in solids}}}}\ (\bibinfo  {publisher} {Springer Science \&
  Business Media},\ \bibinfo {year} {2012})\BibitemShut {NoStop}%
\bibitem [{\citenamefont {Zhou}\ \emph {et~al.}(2020)\citenamefont {Zhou},
  \citenamefont {Choi}, \citenamefont {Choi}, \citenamefont {Landig},
  \citenamefont {Douglas}, \citenamefont {Isoya}, \citenamefont {Jelezko},
  \citenamefont {Onoda}, \citenamefont {Sumiya}, \citenamefont {Cappellaro},
  \citenamefont {Knowles}, \citenamefont {Park},\ and\ \citenamefont
  {Lukin}}]{zhou2020quantum}%
  \BibitemOpen
  \bibfield  {author} {\bibinfo {author} {\bibfnamefont {H.}~\bibnamefont
  {Zhou}}, \bibinfo {author} {\bibfnamefont {J.}~\bibnamefont {Choi}}, \bibinfo
  {author} {\bibfnamefont {S.}~\bibnamefont {Choi}}, \bibinfo {author}
  {\bibfnamefont {R.}~\bibnamefont {Landig}}, \bibinfo {author} {\bibfnamefont
  {A.~M.}\ \bibnamefont {Douglas}}, \bibinfo {author} {\bibfnamefont
  {J.}~\bibnamefont {Isoya}}, \bibinfo {author} {\bibfnamefont
  {F.}~\bibnamefont {Jelezko}}, \bibinfo {author} {\bibfnamefont
  {S.}~\bibnamefont {Onoda}}, \bibinfo {author} {\bibfnamefont
  {H.}~\bibnamefont {Sumiya}}, \bibinfo {author} {\bibfnamefont
  {P.}~\bibnamefont {Cappellaro}}, \bibinfo {author} {\bibfnamefont {H.~S.}\
  \bibnamefont {Knowles}}, \bibinfo {author} {\bibfnamefont {H.}~\bibnamefont
  {Park}}, \ and\ \bibinfo {author} {\bibfnamefont {M.~D.}\ \bibnamefont
  {Lukin}},\ }\href {\doibase 10.1103/PhysRevX.10.031003} {\bibfield  {journal}
  {\bibinfo  {journal} {Physical Review X}\ }\textbf {\bibinfo {volume} {10}},\
  \bibinfo {pages} {031003} (\bibinfo {year} {2020})}\BibitemShut {NoStop}%
\bibitem [{\citenamefont {Choi}\ \emph
  {et~al.}(2017{\natexlab{b}})\citenamefont {Choi}, \citenamefont {Choi},
  \citenamefont {Landig}, \citenamefont {Kucsko}, \citenamefont {Zhou},
  \citenamefont {Isoya}, \citenamefont {Jelezko}, \citenamefont {Onoda},
  \citenamefont {Sumiya}, \citenamefont {Khemani}, \citenamefont {von
  Keyserlingk}, \citenamefont {Yao}, \citenamefont {Demler},\ and\
  \citenamefont {Lukin}}]{choi2017observation}%
  \BibitemOpen
  \bibfield  {author} {\bibinfo {author} {\bibfnamefont {S.}~\bibnamefont
  {Choi}}, \bibinfo {author} {\bibfnamefont {J.}~\bibnamefont {Choi}}, \bibinfo
  {author} {\bibfnamefont {R.}~\bibnamefont {Landig}}, \bibinfo {author}
  {\bibfnamefont {G.}~\bibnamefont {Kucsko}}, \bibinfo {author} {\bibfnamefont
  {H.}~\bibnamefont {Zhou}}, \bibinfo {author} {\bibfnamefont {J.}~\bibnamefont
  {Isoya}}, \bibinfo {author} {\bibfnamefont {F.}~\bibnamefont {Jelezko}},
  \bibinfo {author} {\bibfnamefont {S.}~\bibnamefont {Onoda}}, \bibinfo
  {author} {\bibfnamefont {H.}~\bibnamefont {Sumiya}}, \bibinfo {author}
  {\bibfnamefont {V.}~\bibnamefont {Khemani}}, \bibinfo {author} {\bibfnamefont
  {C.}~\bibnamefont {von Keyserlingk}}, \bibinfo {author} {\bibfnamefont
  {N.~Y.}\ \bibnamefont {Yao}}, \bibinfo {author} {\bibfnamefont
  {E.}~\bibnamefont {Demler}}, \ and\ \bibinfo {author} {\bibfnamefont {M.~D.}\
  \bibnamefont {Lukin}},\ }\href {\doibase 10.1038/nature21426} {\bibfield
  {journal} {\bibinfo  {journal} {Nature}\ }\textbf {\bibinfo {volume} {543}},\
  \bibinfo {pages} {221} (\bibinfo {year} {2017}{\natexlab{b}})}\BibitemShut
  {NoStop}%
\bibitem [{\citenamefont {Zhang}\ \emph {et~al.}(2017)\citenamefont {Zhang},
  \citenamefont {Pagano}, \citenamefont {Hess}, \citenamefont {Kyprianidis},
  \citenamefont {Becker}, \citenamefont {Kaplan}, \citenamefont {Gorshkov},
  \citenamefont {Gong},\ and\ \citenamefont {Monroe}}]{zhang2017observation}%
  \BibitemOpen
  \bibfield  {author} {\bibinfo {author} {\bibfnamefont {J.}~\bibnamefont
  {Zhang}}, \bibinfo {author} {\bibfnamefont {G.}~\bibnamefont {Pagano}},
  \bibinfo {author} {\bibfnamefont {P.~W.}\ \bibnamefont {Hess}}, \bibinfo
  {author} {\bibfnamefont {A.}~\bibnamefont {Kyprianidis}}, \bibinfo {author}
  {\bibfnamefont {P.}~\bibnamefont {Becker}}, \bibinfo {author} {\bibfnamefont
  {H.}~\bibnamefont {Kaplan}}, \bibinfo {author} {\bibfnamefont {A.~V.}\
  \bibnamefont {Gorshkov}}, \bibinfo {author} {\bibfnamefont {Z.~X.}\
  \bibnamefont {Gong}}, \ and\ \bibinfo {author} {\bibfnamefont
  {C.}~\bibnamefont {Monroe}},\ }\href {\doibase 10.1038/nature24654}
  {\bibfield  {journal} {\bibinfo  {journal} {Nature}\ }\textbf {\bibinfo
  {volume} {551}},\ \bibinfo {pages} {601} (\bibinfo {year}
  {2017})}\BibitemShut {NoStop}%
\bibitem [{\citenamefont {Lindner}\ \emph {et~al.}(2011)\citenamefont
  {Lindner}, \citenamefont {Refael},\ and\ \citenamefont
  {Galitski}}]{lindner2011floquet}%
  \BibitemOpen
  \bibfield  {author} {\bibinfo {author} {\bibfnamefont {N.~H.}\ \bibnamefont
  {Lindner}}, \bibinfo {author} {\bibfnamefont {G.}~\bibnamefont {Refael}}, \
  and\ \bibinfo {author} {\bibfnamefont {V.}~\bibnamefont {Galitski}},\ }\href
  {\doibase 10.1038/nphys1926} {\bibfield  {journal} {\bibinfo  {journal}
  {Nature Physics}\ }\textbf {\bibinfo {volume} {7}},\ \bibinfo {pages} {490}
  (\bibinfo {year} {2011})}\BibitemShut {NoStop}%
\bibitem [{\citenamefont {Haeberlen}\ and\ \citenamefont
  {Waugh}(1968)}]{haeberlen1968coherent}%
  \BibitemOpen
  \bibfield  {author} {\bibinfo {author} {\bibfnamefont {U.}~\bibnamefont
  {Haeberlen}}\ and\ \bibinfo {author} {\bibfnamefont {J.~S.}\ \bibnamefont
  {Waugh}},\ }\href {\doibase 10.1103/PhysRev.175.453} {\bibfield  {journal}
  {\bibinfo  {journal} {Physical Review}\ }\textbf {\bibinfo {volume} {175}},\
  \bibinfo {pages} {453} (\bibinfo {year} {1968})}\BibitemShut {NoStop}%
\bibitem [{\citenamefont {Kucsko}\ \emph {et~al.}(2018)\citenamefont {Kucsko},
  \citenamefont {Choi}, \citenamefont {Choi}, \citenamefont {Maurer},
  \citenamefont {Zhou}, \citenamefont {Landig}, \citenamefont {Sumiya},
  \citenamefont {Onoda}, \citenamefont {Isoya}, \citenamefont {Jelezko},
  \citenamefont {Demler}, \citenamefont {Yao},\ and\ \citenamefont
  {Lukin}}]{kucsko2018critical}%
  \BibitemOpen
  \bibfield  {author} {\bibinfo {author} {\bibfnamefont {G.}~\bibnamefont
  {Kucsko}}, \bibinfo {author} {\bibfnamefont {S.}~\bibnamefont {Choi}},
  \bibinfo {author} {\bibfnamefont {J.}~\bibnamefont {Choi}}, \bibinfo {author}
  {\bibfnamefont {P.~C.}\ \bibnamefont {Maurer}}, \bibinfo {author}
  {\bibfnamefont {H.}~\bibnamefont {Zhou}}, \bibinfo {author} {\bibfnamefont
  {R.}~\bibnamefont {Landig}}, \bibinfo {author} {\bibfnamefont
  {H.}~\bibnamefont {Sumiya}}, \bibinfo {author} {\bibfnamefont
  {S.}~\bibnamefont {Onoda}}, \bibinfo {author} {\bibfnamefont
  {J.}~\bibnamefont {Isoya}}, \bibinfo {author} {\bibfnamefont
  {F.}~\bibnamefont {Jelezko}}, \bibinfo {author} {\bibfnamefont
  {E.}~\bibnamefont {Demler}}, \bibinfo {author} {\bibfnamefont {N.~Y.}\
  \bibnamefont {Yao}}, \ and\ \bibinfo {author} {\bibfnamefont {M.~D.}\
  \bibnamefont {Lukin}},\ }\href {\doibase 10.1103/PhysRevLett.121.023601}
  {\bibfield  {journal} {\bibinfo  {journal} {Physical Review Letters}\
  }\textbf {\bibinfo {volume} {121}},\ \bibinfo {pages} {023601} (\bibinfo
  {year} {2018})}\BibitemShut {NoStop}%
\bibitem [{\citenamefont {Cory}\ \emph
  {et~al.}(1990{\natexlab{b}})\citenamefont {Cory}, \citenamefont {Miller},
  \citenamefont {Turner},\ and\ \citenamefont {Garroway}}]{cory1990multiple}%
  \BibitemOpen
  \bibfield  {author} {\bibinfo {author} {\bibfnamefont {D.}~\bibnamefont
  {Cory}}, \bibinfo {author} {\bibfnamefont {J.}~\bibnamefont {Miller}},
  \bibinfo {author} {\bibfnamefont {R.}~\bibnamefont {Turner}}, \ and\ \bibinfo
  {author} {\bibfnamefont {A.}~\bibnamefont {Garroway}},\ }\href {\doibase
  10.1080/00268979000101031} {\bibfield  {journal} {\bibinfo  {journal}
  {Molecular Physics}\ }\textbf {\bibinfo {volume} {70}},\ \bibinfo {pages}
  {331} (\bibinfo {year} {1990}{\natexlab{b}})}\BibitemShut {NoStop}%
\bibitem [{\citenamefont {Mansfield}(1971)}]{mansfield1971symmetrized}%
  \BibitemOpen
  \bibfield  {author} {\bibinfo {author} {\bibfnamefont {P.}~\bibnamefont
  {Mansfield}},\ }\href {\doibase 10.1088/0022-3719/4/11/020} {\bibfield
  {journal} {\bibinfo  {journal} {Journal of Physics C: Solid State Physics}\
  }\textbf {\bibinfo {volume} {4}},\ \bibinfo {pages} {1444} (\bibinfo {year}
  {1971})}\BibitemShut {NoStop}%
\bibitem [{\citenamefont {Hahn}(1950)}]{hahn1950spin}%
  \BibitemOpen
  \bibfield  {author} {\bibinfo {author} {\bibfnamefont {E.~L.}\ \bibnamefont
  {Hahn}},\ }\href {\doibase 10.1103/PhysRev.80.580} {\bibfield  {journal}
  {\bibinfo  {journal} {Physical Review}\ }\textbf {\bibinfo {volume} {80}},\
  \bibinfo {pages} {580} (\bibinfo {year} {1950})}\BibitemShut {NoStop}%
\bibitem [{oth()}]{otherpaper}%
  \BibitemOpen
  \href@noop {} {}\bibinfo {note} {See Accompanying Paper}\BibitemShut
  {NoStop}%
\bibitem [{\citenamefont {Martin}\ \emph {et~al.}(2022)\citenamefont {Martin},
  \citenamefont {Zhou}, \citenamefont {Leitao}, \citenamefont {Maskara},
  \citenamefont {Makarova}, \citenamefont {Gao}, \citenamefont {Zhu},
  \citenamefont {Park}, \citenamefont {Tyler}, \citenamefont {Park},
  \citenamefont {Choi},\ and\ \citenamefont {Lukin}}]{martin2022controlling}%
  \BibitemOpen
  \bibfield  {author} {\bibinfo {author} {\bibfnamefont {L.~S.}\ \bibnamefont
  {Martin}}, \bibinfo {author} {\bibfnamefont {H.}~\bibnamefont {Zhou}},
  \bibinfo {author} {\bibfnamefont {N.~T.}\ \bibnamefont {Leitao}}, \bibinfo
  {author} {\bibfnamefont {N.}~\bibnamefont {Maskara}}, \bibinfo {author}
  {\bibfnamefont {O.}~\bibnamefont {Makarova}}, \bibinfo {author}
  {\bibfnamefont {H.}~\bibnamefont {Gao}}, \bibinfo {author} {\bibfnamefont
  {Q.-Z.}\ \bibnamefont {Zhu}}, \bibinfo {author} {\bibfnamefont
  {M.}~\bibnamefont {Park}}, \bibinfo {author} {\bibfnamefont {M.}~\bibnamefont
  {Tyler}}, \bibinfo {author} {\bibfnamefont {H.}~\bibnamefont {Park}},
  \bibinfo {author} {\bibfnamefont {S.}~\bibnamefont {Choi}}, \ and\ \bibinfo
  {author} {\bibfnamefont {M.~D.}\ \bibnamefont {Lukin}},\ }\href {\doibase
  10.48550/arxiv.2209.09297} {\bibfield  {journal} {\bibinfo  {journal} {arXiv
  preprint arXiv:2209.09297}\ } (\bibinfo {year} {2022}),\
  10.48550/arxiv.2209.09297}\BibitemShut {NoStop}%
\bibitem [{\citenamefont {Zu}\ \emph {et~al.}(2021)\citenamefont {Zu},
  \citenamefont {Machado}, \citenamefont {Ye}, \citenamefont {Choi},
  \citenamefont {Kobrin}, \citenamefont {Mittiga}, \citenamefont {Hsieh},
  \citenamefont {Bhattacharyya}, \citenamefont {Markham}, \citenamefont
  {Twitchen}, \citenamefont {Jarmola}, \citenamefont {Budker}, \citenamefont
  {Laumann}, \citenamefont {Moore},\ and\ \citenamefont
  {Yao}}]{zu2021emergent}%
  \BibitemOpen
  \bibfield  {author} {\bibinfo {author} {\bibfnamefont {C.}~\bibnamefont
  {Zu}}, \bibinfo {author} {\bibfnamefont {F.}~\bibnamefont {Machado}},
  \bibinfo {author} {\bibfnamefont {B.}~\bibnamefont {Ye}}, \bibinfo {author}
  {\bibfnamefont {S.}~\bibnamefont {Choi}}, \bibinfo {author} {\bibfnamefont
  {B.}~\bibnamefont {Kobrin}}, \bibinfo {author} {\bibfnamefont
  {T.}~\bibnamefont {Mittiga}}, \bibinfo {author} {\bibfnamefont
  {S.}~\bibnamefont {Hsieh}}, \bibinfo {author} {\bibfnamefont
  {P.}~\bibnamefont {Bhattacharyya}}, \bibinfo {author} {\bibfnamefont
  {M.}~\bibnamefont {Markham}}, \bibinfo {author} {\bibfnamefont
  {D.}~\bibnamefont {Twitchen}}, \bibinfo {author} {\bibfnamefont
  {A.}~\bibnamefont {Jarmola}}, \bibinfo {author} {\bibfnamefont
  {D.}~\bibnamefont {Budker}}, \bibinfo {author} {\bibfnamefont {C.~R.}\
  \bibnamefont {Laumann}}, \bibinfo {author} {\bibfnamefont {J.~E.}\
  \bibnamefont {Moore}}, \ and\ \bibinfo {author} {\bibfnamefont {N.~Y.}\
  \bibnamefont {Yao}},\ }\href {\doibase 10.1038/s41586-021-03763-1} {\bibfield
   {journal} {\bibinfo  {journal} {Nature}\ }\textbf {\bibinfo {volume}
  {597}},\ \bibinfo {pages} {45} (\bibinfo {year} {2021})}\BibitemShut
  {NoStop}%
\bibitem [{\citenamefont {Merkel}\ \emph {et~al.}(2021)\citenamefont {Merkel},
  \citenamefont {{Cova Fari{\~{n}}a}},\ and\ \citenamefont
  {Reiserer}}]{merkel2021dynamical}%
  \BibitemOpen
  \bibfield  {author} {\bibinfo {author} {\bibfnamefont {B.}~\bibnamefont
  {Merkel}}, \bibinfo {author} {\bibfnamefont {P.}~\bibnamefont {{Cova
  Fari{\~{n}}a}}}, \ and\ \bibinfo {author} {\bibfnamefont {A.}~\bibnamefont
  {Reiserer}},\ }\href {\doibase 10.1103/PhysRevLett.127.030501} {\bibfield
  {journal} {\bibinfo  {journal} {Physical Review Letters}\ }\textbf {\bibinfo
  {volume} {127}},\ \bibinfo {pages} {030501} (\bibinfo {year}
  {2021})}\BibitemShut {NoStop}%
\bibitem [{\citenamefont {Khaneja}\ \emph {et~al.}(2005)\citenamefont
  {Khaneja}, \citenamefont {Reiss}, \citenamefont {Kehlet}, \citenamefont
  {Schulte-Herbr{\"{u}}ggen},\ and\ \citenamefont
  {Glaser}}]{khaneja2005optimal}%
  \BibitemOpen
  \bibfield  {author} {\bibinfo {author} {\bibfnamefont {N.}~\bibnamefont
  {Khaneja}}, \bibinfo {author} {\bibfnamefont {T.}~\bibnamefont {Reiss}},
  \bibinfo {author} {\bibfnamefont {C.}~\bibnamefont {Kehlet}}, \bibinfo
  {author} {\bibfnamefont {T.}~\bibnamefont {Schulte-Herbr{\"{u}}ggen}}, \ and\
  \bibinfo {author} {\bibfnamefont {S.~J.}\ \bibnamefont {Glaser}},\ }\href
  {\doibase 10.1016/j.jmr.2004.11.004} {\bibfield  {journal} {\bibinfo
  {journal} {Journal of Magnetic Resonance}\ }\textbf {\bibinfo {volume}
  {172}},\ \bibinfo {pages} {296} (\bibinfo {year} {2005})}\BibitemShut
  {NoStop}%
\bibitem [{\citenamefont {Peng}\ \emph {et~al.}(2021)\citenamefont {Peng},
  \citenamefont {Huang}, \citenamefont {Yin}, \citenamefont {Joseph},
  \citenamefont {Ramanathan},\ and\ \citenamefont {Cappellaro}}]{peng2021deep}%
  \BibitemOpen
  \bibfield  {author} {\bibinfo {author} {\bibfnamefont {P.}~\bibnamefont
  {Peng}}, \bibinfo {author} {\bibfnamefont {X.}~\bibnamefont {Huang}},
  \bibinfo {author} {\bibfnamefont {C.}~\bibnamefont {Yin}}, \bibinfo {author}
  {\bibfnamefont {L.}~\bibnamefont {Joseph}}, \bibinfo {author} {\bibfnamefont
  {C.}~\bibnamefont {Ramanathan}}, \ and\ \bibinfo {author} {\bibfnamefont
  {P.}~\bibnamefont {Cappellaro}},\ }\href {\doibase 10.48550/arxiv.2102.13161}
  {\bibfield  {journal} {\bibinfo  {journal} {arXiv preprint arXiv:2102.13161}\
  } (\bibinfo {year} {2021}),\ 10.48550/arxiv.2102.13161}\BibitemShut {NoStop}%
\bibitem [{\citenamefont {Hohwy}\ and\ \citenamefont
  {Nielsen}(1997)}]{hohwy1997elimination}%
  \BibitemOpen
  \bibfield  {author} {\bibinfo {author} {\bibfnamefont {M.}~\bibnamefont
  {Hohwy}}\ and\ \bibinfo {author} {\bibfnamefont {N.~C.}\ \bibnamefont
  {Nielsen}},\ }\href {\doibase 10.1063/1.473760} {\bibfield  {journal}
  {\bibinfo  {journal} {The Journal of Chemical Physics}\ }\textbf {\bibinfo
  {volume} {106}},\ \bibinfo {pages} {7571} (\bibinfo {year}
  {1997})}\BibitemShut {NoStop}%
\bibitem [{\citenamefont {Mote}\ \emph {et~al.}(2016)\citenamefont {Mote},
  \citenamefont {Agarwal},\ and\ \citenamefont {Madhu}}]{mote2016five}%
  \BibitemOpen
  \bibfield  {author} {\bibinfo {author} {\bibfnamefont {K.~R.}\ \bibnamefont
  {Mote}}, \bibinfo {author} {\bibfnamefont {V.}~\bibnamefont {Agarwal}}, \
  and\ \bibinfo {author} {\bibfnamefont {P.}~\bibnamefont {Madhu}},\ }\href
  {\doibase 10.1016/J.PNMRS.2016.08.001} {\bibfield  {journal} {\bibinfo
  {journal} {Progress in Nuclear Magnetic Resonance Spectroscopy}\ }\textbf
  {\bibinfo {volume} {97}},\ \bibinfo {pages} {1} (\bibinfo {year}
  {2016})}\BibitemShut {NoStop}%
\bibitem [{\citenamefont {Mananga}(2016)}]{mananga2016on}%
  \BibitemOpen
  \bibfield  {author} {\bibinfo {author} {\bibfnamefont {E.~S.}\ \bibnamefont
  {Mananga}},\ }\href {\doibase 10.1016/J.PHYSREP.2015.10.006} {\bibfield
  {journal} {\bibinfo  {journal} {Physics Reports}\ }\textbf {\bibinfo {volume}
  {608}},\ \bibinfo {pages} {1} (\bibinfo {year} {2016})}\BibitemShut {NoStop}%
\bibitem [{zho()}]{zhou2023robust}%
  \BibitemOpen
  \href@noop {} {}\bibinfo {note} {H. Zhou et al., in preparation}\BibitemShut
  {NoStop}%
\bibitem [{lei()}]{leitao2022qudit}%
  \BibitemOpen
  \href@noop {} {}\bibinfo {note} {N. Leitao et al., in
  preparation}\BibitemShut {NoStop}%
\end{thebibliography}%

\newpage
\onecolumngrid
\appendix
\section{Conventions}
\label{sec:Conventions}
See Tab.~\ref{tab:conventions} for a summary of the conventions employed in this manuscript.
\begin{table*}
\begin{tabular}{|c|l|}
    \hline
    $T$ & Floquet period\\\hline
    $t_k$ & Midpoint time of a the $k$-th free evolution period\\\hline
    $\tau_k$ & Duration of the $k$-th free evolution period\\\hline
    $\tau_p$ & Duration of $\pi/2$ pulses\\\hline
    $H_{dip}$& Dipole Hamiltonian\\\hline
    $\tilde{H}(t)$ & Interaction picture Hamiltonian\\\hline
    $\tilde H_{k}$& Interaction picture Hamiltonian during the $k$-th free evolution period\\\hline
    $H_{\textit{eff}}$& Time-independent effective Hamiltonian\\\hline
    $H^{(k)}$& $k$-th order Magnus term for the effective Hamiltonian\\\hline
    $h_i$ & On-site disorder strength for spin $i$\\\hline
    $J_{i,j}$ & Interaction strength between spins $i$ and $j$ \\\hline
    $P_k$& $k$-th global spin rotation pulse\\\hline
    $U_c(t)$&Single spin rotation unitary due to the control field\\\hline
    $U(T)$& Unitary operator for the evolution over one full Floquet cycle\\\hline
    $\tilde{S}^z(t)$& Interaction picture $S^z$ vector at time $t$ under an ideal sequence\\\hline 
    $S^{\mu}$&Spin basis (e.g. Pauli spin operators)\\\hline
    $F_{\mu}(t)$& The $S^{\mu}$ coefficient of $\tilde S^{z}$ at time $t$\\\hline
    $F_{\mu,k}$ & The $S^{\mu}$ coefficient of $\tilde {S}^{z}$ during free evolution frame $k$\\\hline
    $F_{<k}^\mu$& Accumulated disorder through pulse $k$, $\sum_{j<k}F_{\mu,j}(\tau_j+\frac4\pi\tau_p)$\\\hline
    $F_{>k}^\mu$& Accumulated disorder after pulse $k$, $\sum_{j>k}F_{\mu,j}(\tau_j+\frac4\pi\tau_p)$\\\hline
    $I_{<k}^\mu$& Accumulated Ising interaction through pulse $k$, $\sum_{j<k}|F_{\mu,j}|(\tau_j+\tau_p)$\\\hline
    $I_{>k}^\mu$& Accumulated Ising interaction after pulse $ k$, $\sum_{j>k}|F_{\mu,j}|(\tau_j+\tau_p)$\\\hline
    $F^{\nu,\rho}_{<k}$& Accumulated first-order disorder-disorder effect through pulse $k$, $\sum_{l=1}^kF_{\nu,l}(\tau_l+\frac4\pi\tau_p)(F^\rho_{<l}-F^\rho_{>l})$\\\hline
    $\overline{F}^\mu$& Total zeroth-order disorder effect, $\sum_{k=1}^{n}F_{\mu,k}(\tau_k+\frac4\pi\tau_p)$\\\hline
    $\overline{I}^\mu$& Total zeroth-order Ising interaction, $\sum_{k=1}^{n}|F_{\mu,k}|(\tau_k+\tau_p)$\\\hline
    $\overline{F}^{\nu,\rho}$& Total first-order disorder-disorder effect $\sum_{k=1}^{n}F_{\nu,k}(\tau_k+\frac4\pi\tau_p)F_{<k}^\rho$\\\hline
    $\Op^\alpha$ & Operator basis for interaction-picture Hamiltonian \\\hline
    $c_\alpha(t)$ & Time-dependent coefficients of $\Op^\alpha$ in the Hamiltonian \\\hline
    $\cc_{\alpha}(t)$& Accumulation of $c_{\alpha}(t_1)$ over $t_1\in[0,t]$\\\hline
    $c_{\mu,k}$ & Discrete version of $c_\mu(t)$ in the $k$th free evolution frame \\\hline
    $\cpl_{\mu,k}(\theta)$ & $c_\mu(t)$ during the pulse preceding the $k$th free evolution time, as a function of the rotation angle $\theta\in[0,\pi/2]$ \\\hline
    $\cm_{\mu,k}(\theta)$ & $c_\mu(t)$ during the pulse following the $k$th free evolution time, as a function of the rotation angle $\theta\in[0,\pi/2]$ \\\hline
    $r$ & Angular rotation rate during a pulse \\\hline
    $C_{\mu,k}$ & Total accumulation of $c_{\mu}(t)$ over $t\in[t_k-\tau_{k}/2-\tau_p,t_k+\tau_k/2+\tau_p]$ \\\hline
    $\cc_{\mu,k}$ & Accumulated coefficient, $\sum_{l<k}C_{\mu,l}$\\\hline
    $q_{\alpha,k,k+1}(\theta)$ & Cross terms during pulses that involve both preceding and following frames $k$ and $k+1$ \\\hline
    $A_{free}$ & First-order contribution from free evolution frames\\\hline
    $C^{(1)}_{\alpha,\beta,k}$ & Frame $k$'s first order effect with itself\\\hline
    $P_{k,k+1,\alpha,\beta}$ & Finite pulse correction for $k$th pulse\\\hline
    %$g_{\alpha,k,k+1}(\theta)$ & total coefficient during finite pulse \\\hline
    %$Q_{\alpha,a,b}$ & the cross term affect coming from all pulses between frames $a$ and $b$\\\hline

\end{tabular}\\
\caption{Summary of conventions employed in this paper.\label{tab:conventions}}
\end{table*}

\section{Derivation of First-Order Magnus Formalism}
\label{supp:full1storder}

To develop the full expression at first order, we extend the formalism developed in Ref.~\cite{choi2020robust}. To keep the expressions fully general, we do not restrict to a specific qubit Hamiltonian here, and specialize to the dipolar Hamiltonian only in the following sections.

Following Ref.~\cite{choi2020robust}, we can separate the evolution into free evolution periods and evolution during pulses.
We will write the coefficient of a given operator $\Op^\alpha$ during the $k$-th free evolution period as $c_{\alpha}(t)=c_{\alpha,k}$, and during the $\pi/2$ pulse after the $k$-th free evolution period as
\begin{align}
c_\alpha(t)=\cm_{\alpha,k}\qty(\frac tr)+q_{\alpha,k,k+1}\qty(\frac tr)+\cpl_{\alpha,k+1}\qty(\frac tr).
\end{align}
Here, $r$ is the rate of angular precession under the applied pulses, and the rotation angles during the $\pi/2$ pulse are given by $\theta=t/r$.
The first term $\cm_{\alpha,k}(\theta)$ describes the finite pulse duration contribution from the $k$-th frame that precedes the pulse, while $\cpl_{\alpha,k+1}(\theta)$ describes the contribution from the ($k+1$)-th frame that follows the pulse.
$q_{\alpha,k,k+1}(\theta)$ is an additional cross-term between the two frames that arises for certain types of interaction terms.
Note that similar to Ref.~\cite{choi2020robust}, in our pulse sequence composed of $\pi/2$ pulses and $\pi$ pulses, we treat each $\pi$ pulse as a combination of two $\pi/2$ pulses with zero free evolution time in between.

As a concrete example to illustrate these terms, let us consider a rotation that transformed the $S^z$ operator into $S^x$, i.e. $\tilde{S}^z(\theta)=\cos\theta S^z+\sin\theta S^x$.
For an Ising interaction $H_I=JS_i^zS_j^z$, the time-dependent operator would be
\begin{align}
\label{eq:isingrotation}
\tilde{H}_I(t)&=J\qty[\cos^2\theta S_i^zS_j^z+\sin\theta\cos\theta(S_i^xS_j^z+S_i^zS_j^x)+\sin^2\theta S_i^xS_j^x].
\end{align}
The three terms in the parenthesis correspond to the $\cm_{\alpha,k}(\theta)$, $q_{\alpha,k,k+1}(\theta)$ and $\cpl_{\alpha,k+1}(\theta)$ terms, respectively.

With this representation in hand, we proceed by rewriting the integral in Eq.~(\ref{eq:firstordermagnusinitial}) as a summation over the distinct blocks.
Let us examine the first term, which integrates all $c_\beta$ terms occurring temporally before $c_\alpha$:
\begin{align}
\label{eq:main_first_order}
A=\int_0^Tdt_1c_\alpha(t_1)\int_0^{t_1}dt_2c_\beta(t_2)
\end{align}

To compute this, we first define the integral of the coefficient of a given frame, including its finite pulse duration effects:
\begin{align}
C_{\alpha,k}=r\int_{0}^{\pi/2}\cpl_{\alpha,k}(\theta)d\theta+\int_{0}^{\tau_k}c_{\alpha,k}dt+r\int_{0}^{\pi/2}\cm_{\alpha,k}(\theta)d\theta.
\end{align}
This can be viewed as a simple extension of the effective free evolution time.

\begin{figure}
\begin{center}
\includegraphics[width=0.6\columnwidth]{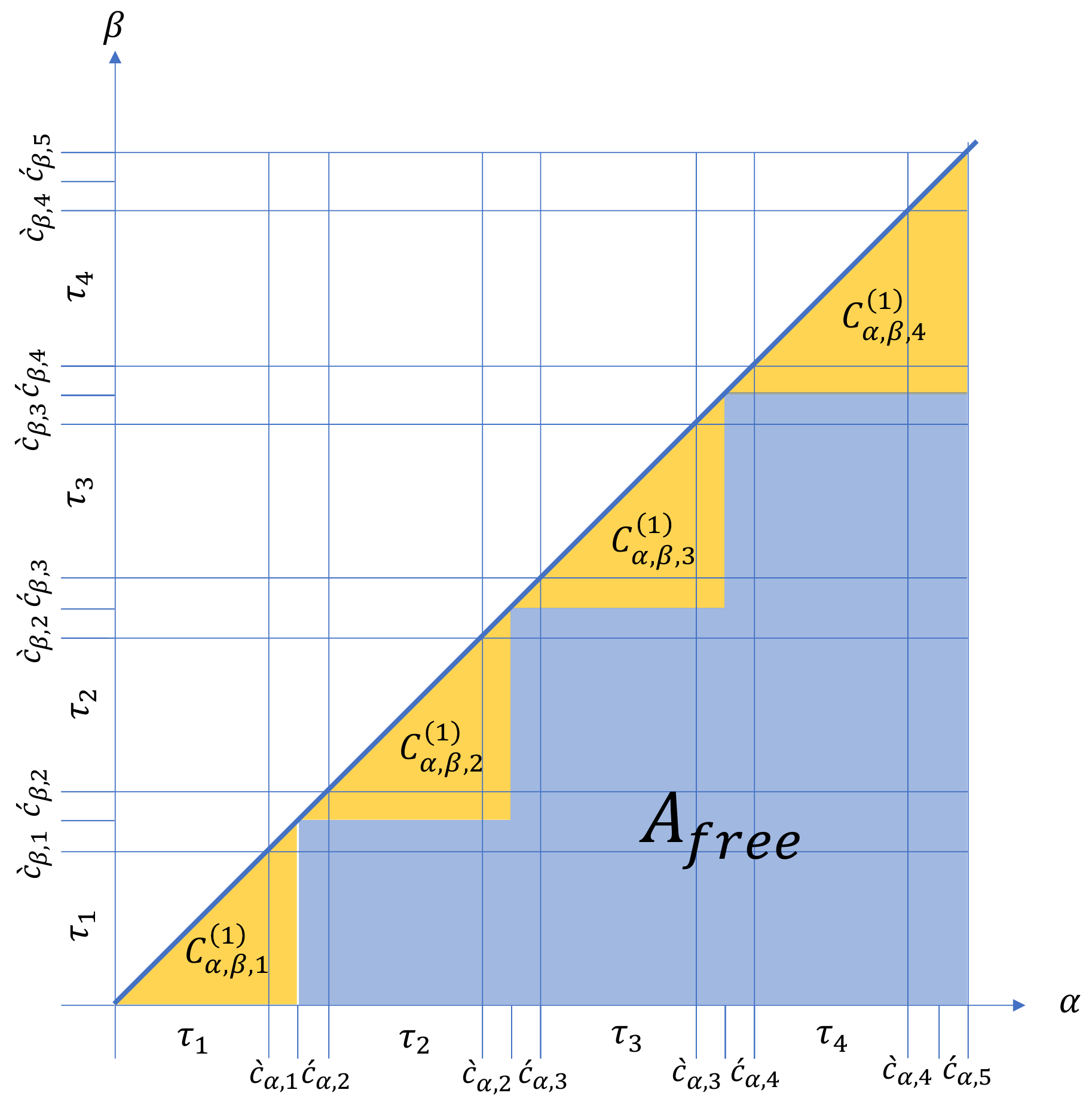}
\caption{{\bf Visualization of first-order cancellation rule derivation, in the absence of $q$-terms.} The illustration shows the different components of the first-order expression. The lower triangle of the grid is the first-order integral. In reality, the rising and falling edges of neighboring frames will overlap, i.e. $\cm_{\alpha,1}$ is not completely before $\cpl_{\alpha,2}$, which leads to the additional term $P_{k,k+1,\alpha,\beta}$.}
\label{fig:first_order_intuition}
\end{center}
\end{figure}

We can then decompose the inner integral in Eq.~(\ref{eq:main_first_order}) into three parts (ignoring additional contributions from $q_{\alpha,k,k+1}$-terms for now): a contribution from previous, non-overlapping free evolution times, together with their surrounding pulses ($A_{free}$); a contribution from integrating both time variables within the same free evolution period, corresponding to first-order contributions within the same frame ($C_{\alpha,\beta,k}^{(1)}$); and a further correction arising from the pulse overlaps of neighboring free evolution times ($P_{k,k+1,\alpha,\beta}$).
This is illustrated in Fig.~\ref{fig:first_order_intuition}.

More concretely, the first term describes contributions where $t_1$ lies within the $k$-th frame, and $t_2$ originates from an earlier frame. As most of these contributions will be temporally non-overlapping, we can factorize these contributions as
\begin{align}
\label{eq:Afree}
A_{free}=\sum_{k=1}^n C_{\alpha,k}\sum_{j=1}^{k-1}C_{\beta,j}.
\end{align}

The next term describes contributions where both $t_1$ and $t_2$ come from the $k$-th frame, with $t_2<t_1\in[t_{k-1}+\frac{\tau_{k-1}}2,t_{k+1}-\frac{\tau_{k+1}}2]$, i.e. the first-order Magnus contribution of a frame with itself.
We can explicitly write this as
\begin{align}
\label{eq:C1}
C^{(1)}_{\alpha,\beta,k}=\int_{t_{k-1}+\frac{\tau_{k-1}}2}^{t_{k+1}-\frac{\tau_{k+1}}2}c_{\alpha}(t_1)dt_1\int_{t_{k-1}+\frac{\tau_{k-1}}2}^{t_1}c_{\beta}(t_2)dt_2.
\end{align}
Note that this term is usually zero in our case, as the commutator $\qty[\Op^\alpha,\Op^\beta]$ vanishes when $\alpha=\beta$, and otherwise $c_\alpha$, $c_\beta$ are both non-zero within the same free evolution period only when they originate from different types of non-commuting Hamiltonians, e.g. one coming from local $S^z$ disorder, and the other coming from Heisenberg interactions.

Finally, we have additional corrections $P_{k,k+1,\alpha,\beta}$ that arise from the overlap in terms due to the pulses: for the $k$-th frame, we overcounted the overlap contribution with the previous $(k-1)$-th frame by assuming that the $k$-th frame came completely after the $(k-1)$-th frame, but undercounted the overlap contribution with the next $(k+1)$-th frame.
Explicit calculation shows that this results in a first order pulse correction that is related to both the preceding and subsequent frame:
\begin{align}
\label{eq:pairproduct}
P_{k,k+1,\alpha,\beta}=\int_0^{\pi/2}\cm_{\alpha,k}(\theta_1)rd\theta_1\int_{0}^{\theta_1}\cpl_{\beta,k+1}(\theta_2)rd\theta_2-\int_{0}^{\pi/2}\cpl_{\alpha,k+1}(\theta_1)rd\theta_1\int_{\theta_1}^{\pi/2}\cm_{\beta,k}(\theta_2)rd\theta_2.
\end{align}

Thus, neglecting all terms that directly dependent on multiple pulses (see $q$-terms below), we can express the integral in a clean manner as
\begin{align}
 A=A_{free}+\sum_{k=1}^{n-1}P_{k,k+1,\alpha,\beta}+\sum_{k=1}^{n}C^{(1)}_{\alpha,\beta,k}.
\end{align}

For certain interaction terms such as the Ising Hamiltonian, there is an additional contribution we need to keep track of, the $q_{\alpha,k,k+1}(\theta)$ terms described in Eq.~(\ref{eq:isingrotation}).
These terms come from the fact that the Ising interaction transforms as the square of the frame coefficients, introducing additional cross-terms when expanding the square.
These terms are ignored in the main text, as they are negligible for our disorder-dominated system, but we will analyze them in more detail here.

We can perform a similar decomposition of the terms as above, now adding in the contributions from the $q$-terms.
We can treat the $q$-terms as a special type of free evolution frame, and decompose the sum into the three types again, this time keeping track also of whether the other term is a $q$-term or a regular free evolution period.

Similar to Eq.~(\ref{eq:Afree}), we can evaluate the first-order contributions involving a single $q$-term and a single free evolution frame as
\begin{align}
QA_{free}=\sum_{k=1}^nC_{\alpha,k}\sum_{j=1}^{k-1}Q_{\beta,j,j+1}+\sum_{k=1}^nQ_{\alpha,k,k+1}\sum_{j=1}^{k}C_{\beta,j},
\end{align}
where
\begin{align}
Q_{\alpha,k,k+1}=\int_0^{\pi/2}q_{\alpha,k,k+1}(\theta)d\theta
\end{align}
is the integral of the $q$-term during a given pulse.

In analogy to the corrections $P_{k,k+1,\alpha,\beta}$ found above, we also have similar corrections here
\begin{align}
QP_{k,k+1,\alpha,\beta}&=\int_0^{\pi/2}q_{\alpha,k,k+1}(\theta_1)rd\theta_1\int_0^{\theta_1}\cpl_{\beta,k+1}(\theta_2)rd\theta_2-\int_{0}^{\pi/2}q_{\alpha,k,k+1}(\theta_1)rd\theta_1\int_{\theta_1}^{\pi/2}\cm_{\beta,k}(\theta_2)rd\theta_2\nonumber\\
&+\int_0^{\pi/2}\cm_{\alpha,k}(\theta_1)rd\theta_1\int_0^{\theta_1}q_{\beta,k,k+1}(\theta_2)rd\theta_2-\int_0^{\pi/2}\cpl_{\alpha,k}(\theta_1)rd\theta_1\int_{\theta_1}^{\pi/2}q_{\beta,k-1,k}(\theta_2)rd\theta_2.
\end{align}

Finally, we also have corrections coming from the first-order contributions between $q$-terms at different times and in the same pulse
\begin{align}
Q_{self}=\sum_{k=1}^{n}Q_{\alpha,k,k+1}\sum_{l=1}^{k-1}Q_{\beta, l,l+1}+\sum_{k=1}^{n}\int_0^{\pi/2}q_{\alpha,k,k+1}(\theta_1)rd\theta_1\int_{\theta_1}^{\pi/2}q_{\beta,k,k+1}(\theta_2)rd\theta_2.
\end{align}

Putting all of this together, the final, complete expression for first-order terms is
\begin{align}
A=A_{free}+\sum_{k=1}^{n-1}P_{k,k+1,\alpha,\beta}+\sum_{k=1}^{n}C^{(1)}_{\alpha,\beta,k}+QA_{free}+Q_{self}+\sum_{k=1}^nQP_{k,k+1,\alpha,\beta}
\label{eq:complete_first_order}
\end{align}

\section{Derivation of First-Order Cancellation Rules}
\label{supp:full1storderrules}

We will now apply the preceding general calculations to specific first-order terms, in order to derive first-order decoupling rules.
As we shall see, in many cases of interest, a lot of the terms in Eq.~(\ref{eq:complete_first_order}) will drop out, resulting in simple expressions.

\subsection{Disorder-Disorder Rules}
Let us start with first-order disorder-disorder contributions, involving commutators between disorder at different times.
Since this Hamiltonian involves only single-qubit terms, there will be no $q$-terms.
Furthermore, there are no $C^{(1)}_{\alpha,\beta,k}$ terms, as the operator in each frame commutes with itself.
We thus have
\begin{align}
A_{dis-dis}=A_{free}+\sum_{k=1}^{n-1}P_{k,k+1,\alpha,\beta}.
\end{align}

Examining the transformation of the operators for different frames, we have
\begin{align}
c_{\alpha,k}&\to F_{\mu,k},\\
\cpl_{\alpha,k}(\theta)&\to F_{\mu,k}\sin(\theta),\\
\cm_{\alpha,k}(\theta)&\to F_{\mu,k}\cos(\theta),\\
C_{\alpha,k}&= F_{\mu,k}\qty(\tau_k+\frac{4\tau_p}{\pi}).
\end{align}

Plugging this into the preceding definitions of the individual terms, we find
\begin{align}
A_{free}&=\sum_{k=1}^nF_{\mu,k}\qty(\tau_k+\frac{4\tau_p}{\pi})\sum_{l=1}^{k-1}F_{\nu,l}\qty(\tau_l+\frac{4\tau_p}{\pi}),\\
P_{k,k+1,dis,dis}&= \qty(F_{\mu,k}F_{\nu,k+1}-F_{\mu,k+1}F_{\nu,k})\qty(1-\frac\pi4)\qty(\frac{2\tau_p}{\pi})^2,
\end{align}

Further plugging this into the full expression Eq.~(\ref{eq:firstordermagnusinitial}) for the first-order disorder-disorder term, we arrive at the full expression for the main term
\begin{align}
    \sum_{k=1}^nF_{\mu,k}(\tau_k+\frac4\pi\tau_p)F^\nu_{<k}-\frac12\bar F^\mu \bar F^\nu,
\end{align}
and the finite pulse correction
\begin{align}
    \left(1-\frac\pi4\right)\left(\frac{2\tau_p}{\pi}\right)\sum_{k=1}^nF_{\mu,k}F_{\nu,k+1}-F_{\nu,k}F_{\mu,k+1},
\end{align}
as described in Tab.~\ref{tab:FirstOrder}.

Based on these expressions, we can formulate relatively simple rules for their cancellation in sequence design.
The expression $P_{k,k+1,\alpha,\beta}$ involves a term that can be rewritten as $\vec{F}_k\times \vec{F}_{k+1}$, and thus has the same conditions for cancellation as zeroth-order rotation angle errors~\cite{choi2020robust}.

Due to the rapid spin echo structure found in many decoupling sequences for disorder-dominated systems, e.g. DROID-60 in Ref.~\cite{zhou2020quantum}, the majority of terms in $A_{free}$ are also cancelled in the inner sum, and the only contribution remaining is from the commutator between a spin echo pair and the intermediate pulse frame that the $\pi$ pulse uses.
To give a concrete example of this remaining contribution, consider a sequence of two $\pi$ pulses around $X$, which implements the following frame transformations $+Z \rightarrow +Y \rightarrow -Z \rightarrow -Y$, with $+Z$ and $-Z$ being longer free evolution frames, and $+Y$ and $-Y$ being shorter frames with zero free evolution time and only pulse effects.
The first-order contribution from this will then be proportional to the commutator between $Z$ and $Y$, and changes sign both when we flip the sign of one of the operators (e.g. $+Z \rightarrow -Y \rightarrow -Z \rightarrow +Y$), as well as when we switch the order of the operators (e.g. $+Y \rightarrow +Z \rightarrow -Y \rightarrow -Z$).
Thus, this term has the same transformation properties as a rotation angle error that acts only within such spin echo blocks.

\subsection{Disorder-Heisenberg Rules}
The next term we consider is the first-order disorder-Heisenberg contribution, which was the dominant imperfection in the previous DROID-60 sequence~\cite{zhou2020quantum} and key to the design of improved sensing sequences such as DIRAC2~\cite{otherpaper}.

As shown in Eq.~(\ref{eq:firstordermagnusinitial}), we can choose an index ordering where disorder is after Heisenberg interactions, such that $\alpha$ is a disorder index and $\beta$ is a Heisenberg interaction index.
The case where both are Heisenberg indices gives zero contribution, as the operator terms are equal to the fixed Heisenberg Hamiltonian and hence commute.
As the Heisenberg interaction is invariant under frame transformations, the coefficients can be chosen to take a particularly simple form:
\begin{align}
    c_{\beta,k}&\to 1,
    \\
    \cpl_{\beta,k}(\theta)&\to 1,
    \\
    \cm_{\beta,k}(\theta)&\to 0,
    \\
    C_{\beta,k}&=\tau_k+\tau_p.
\end{align}
Plugging these into the preceding expressions, we find
\begin{align}
A_{free}&=\sum_{k=1}^nF_{\mu,k}\qty(\tau_k+\frac{4\tau_p}{\pi})\sum_{j=1}^{k-1}\qty(\tau_j+\tau_p),\\
P_{k,k+1,\alpha,\beta}&=F_{\mu,k}\qty(\frac{\pi}{2}-1)\qty(\frac{2\tau_p}{\pi})^2,\\
C_{\alpha,\beta,k}^{(1)}&=F_{\mu,k}\left[\int_0^{\pi/2}\sin(\theta_1)rd\theta_1\int_0^{\theta_1}rd\theta_2+\int_0^{\tau_k}dt_1\qty(\tau_p+\int_0^{t_1}dt_2)+\int_0^{\pi/2}\cos(\theta)rd\theta\qty(\tau_p+\tau_k)\right]\nonumber\\
&=F_{\mu,k}\left[\qty(\frac{2\tau_p}{\pi})^2+\tau_k\tau_p+\frac{\tau_k^2}{2}+(\tau_p+\tau_k)\qty(\frac{2\tau_p}{\pi})\right].
\end{align}

We can simplify the sum of the last two contributions
\begin{align}
P_{k,k+1,\alpha,\beta}+C^{(1)}_{\alpha,\beta,k}&=F_{\mu,k}\left(\tau_k\tau_p+\frac12\tau_k^2+\qty(2\tau_p+\tau_k)\qty(\frac{2\tau_p}{\pi})\right)\nonumber\\
&=F_{\mu,k}\qty(\tau_p+\frac12\tau_k)\qty(\tau_k+\frac{4\tau_p}{\pi}).
\end{align}

Adding the corrections together, we get
\begin{align}
A=\sum_{k=1}^{n}F_{\mu,k}(\tau_k+\frac{4\tau_p}{\pi})\sum_{j=1}^{k-1}(\tau_j+\tau_p+\tau_p+\frac12{\tau_k})=\sum_{k=1}^{n}F_{\mu,k}t_k(\tau_k+\frac{4\tau_p}{\pi}),
\end{align}
where $t_k$ is the midpoint of the $k$th free evolution frame.
The remaining term in Eq.~(\ref{eq:firstordermagnusinitial}) can be evaluated to be
\begin{align}
\frac12\cc_{\alpha}(T)\cc_{\beta}(T)=\bar F^\mu \frac{T}{2},
\end{align}
which combined give us the full algebraic condition for first-order disorder-Heisenberg decoupling
\begin{align}
\sum_{k=1}^nF_{\mu,k}\left(\tau_k+\frac4\pi\tau_p\right)\left(t_k-\frac T2\right).
\end{align}

As described in the main text and in Ref.~\cite{otherpaper}, there is a relatively simple intuition for these contributions, which we visualize using dipole balancing.
If we associate a charge to each frame, with $+1$($-1$) values of $F_{\mu,k}$ being a positive(negative) charge, then the above expression corresponds to the product of charges ($F_{\mu,k}$) with their center-of-mass location $(t_k-T/2)$, which is precisely the definition of a dipole.
Thus, geometrically, we can visualize the cancellation of first-order disorder-Heisenberg contributions as requiring that the net dipole corresponding to a frame configuration to be 0.

\subsection{Disorder-Ising Rules}
Next we move on to the Ising contributions, starting with first-order disorder-Ising terms.
For this, we use $c_{\alpha,k}$ from the disorder term, and we use the following for the $c_{\beta,k}$ terms:
\begin{align}
c_{\beta,k}&\to |F_{\nu,k}|, % or F_{\nu,k}^2
\\
\cpl_{\beta,k}(\theta)&\to |F_{\nu,k}|\sin^2(\theta),
\\
\cm_{\beta,k}(\theta)&\to |F_{\nu,k}|\cos^2(\theta),
\\
q_{\beta,k,k+1}(\theta)&\to F_{\nu,k}F_{\rho,k+1}\sin(\theta)\cos(\theta),
\\
C_{\beta,k}&=|F_{\nu,k}|(\tau_k+\tau_p),
\\
Q_{\beta,k,k+1}&=F_{\nu,k}F_{\rho,k+1}\frac{\tau_p}{\pi}.
\end{align}

Plugging these into the definitions for the individual terms, we find

\begin{align}
A_{free}&=\sum_{k=1}^{n}F_{\mu,k}\left(\tau_k+\frac{4\tau_p}{\pi}\right)\sum_{l=1}^{k-1}|F_{\nu,l}|\left(\tau_l+\tau_p\right),
\\
P_{k,k+1,dis,isi}&=\left(F_{\mu,k}|F_{\nu,k+1}|-F_{\mu,k+1}|F_{\nu,k}|\right)\left(\frac{\pi}{4}-\frac23\right)\left(\frac{2\tau_p}{\pi}\right)^2.
\end{align}

The term $C^{(1)}_{\alpha,\beta,k}$ will not contribute, as the disorder and Ising Hamiltonian within the same free evolution time commute with each other, $[S^\mu\otimes I, S^\mu \otimes S^\mu]=0$.

The algebraic conditions in the Tab.~\ref{tab:FirstOrder} are based on the preceding expressions, and ignore the $q$-terms.
Combining $A_{free}$ with the rest of the terms gives the main term:
\begin{align}
    \sum_{k=1}^nF_{\mu,k}\left(\tau_k+\frac4\pi\tau_p\right)I_{<k}^\nu-\frac12\bar F^\mu\bar I^\nu.
\end{align}
Summing over pulses in the pulse term gives the finite pulse correction
\begin{align}
    \sum_{k=1}^n\left(\frac{2\tau_p}{\pi}\right)^2\left(\frac\pi4-\frac23\right)\left(F_{\mu,k}|F_{\nu,k+1}|-|F_{\nu,k}|F_{\mu,k+1}\right).
\end{align}
The main term will vanish with the same fast spin echo blocks as that found in the first-order disorder-disorder term, and the finite pulse correction can be viewed as a simple generalization of, e.g. the chirality condition in Ref.~\cite{choi2020robust}.

We now further evaluate the $q$-terms.
As any two adjacent frames will have different operators due to the frame change, we will have no contribution when $\nu=\rho$. Explicitly plugging into the above expressions gives
\begin{align}
QA_{free}&=\sum_{k=1}^nF_{\alpha,k}(\tau_k+\frac4\pi\tau_p)\sum_{j=1}^{k-1}F_{\nu,j}F_{\rho,j+1}\frac{\tau_p}{\pi},
\\
QP_{k,k+1,dis,isi}&=\frac16\left(F_{\mu,k}F_{\nu,k}F_{\rho,k+1}- F_{\mu,k}F_{\nu,k-1}F_{\rho,k}\right)\left(\frac{2\tau_p}{\pi}\right)^2,
\\
Q_{self}&=0.
\end{align}

\subsection{Ising-Ising Rules}
We will now compute the first-order Ising-Ising term. Using the definitions of the individual terms as in the previous calculation (taking both $\alpha$ and $\beta$ to be Ising indices), we find
\begin{align}
A_{free}&=\sum_{k=1}^n|F_{\mu,k}|\left(\tau_k+\tau_p\right)\sum_{l=1}^{k-1}|F_{\nu,l}|(\tau_l+\tau_p)\nonumber\\
&=\sum_{k=1}^n|F_{\mu,k}|\left(\tau_k+\tau_p\right)I_{<a}^\nu,\\
P_{k,k+1,isi,isi}&=\left(|F_{\mu,k}||F_{\nu,k+1}|-|F_{\mu,k+1}||F_{\nu,k}|\right)\left(\frac{\pi^2}{32}-\frac14\right)\left(\frac{2\tau_p}{\pi}\right)^2
\end{align}
Similar to the disorder-disorder case and the disorder-Ising case, we have no contribution from the $C^{(1)}_{\alpha,\beta,k}$ terms. Next, we calculate the contribution from the $q$-terms. Directly plugging things in
\begin{align}
QA_{free}&=\sum_{k=1}^n|F_{\mu,k}|(\tau_k+\tau_p)\sum_{j=1}^{k-1}F_{\nu,k}F_{\rho,k+1}\frac{\tau_p}{\pi}+\sum_{k=1}^nF_{\mu,k}F_{\rho,k+1}\frac{\tau_p}{\pi}\sum_{l=1}^k|F_{\nu,l}|(\tau_l+\tau_p),\\
QP_{k,k+1,isi,isi}&=\frac{\tau_p^2}{8\pi}\left(F_{\mu,k}F_{\lambda,k+1}|F_{\nu,k+1}|-F_{\mu,k}F_{\lambda,k+1}|F_{\nu,k}|+|F_{\mu,k}|F_{\nu,k}F_{\rho,k+1}-|F_{\mu,k}|F_{\nu,k-1}F_{\rho,k}\right),\\
Q_{self}&=\sum_{k=1}^nF_{\mu,k}F_{\lambda,k+1}\frac{\tau_p}{\pi}\sum_{l=1}^{k-1}F_{\nu,l}F_{\rho,l+1}\frac{\tau_p}{\pi}+\sum_{k=1}^{n}\frac{\tau_p^2}{2\pi^2}F_{\mu,k}F_{\lambda,k+1}F_{\nu,k}F_{\rho,k+1}.
\end{align}

\subsection{Ising-Heisenberg Rules}
Finally, we calculate the first-order Ising-Heisenberg terms.
Taking $\alpha$ to be the Ising terms, we can repeat the derivation from disorder-Heisenberg rules to find that the main terms can be written as
\begin{align}
\sum_{k=1}^n|F_{\mu,k}|(\tau_k+\tau_p)(t_k-\frac{T}2),
\end{align}
which simply replaces the $F_{\mu,k}$ in the original disorder-Heisenberg rule by $|F_{\mu,k}|$.

Although this term is no longer cancelled by imposing a zero net dipole, we can still formulate a simple condition for it to be cancelled: if we have ``balanced" rows, in which the center of mass of each row is in the middle, then this term will be cancelled.
For example, a simple mirror symmetrization will cancel this first-order term.

For the $q$-terms, using the expressions found above, we can easily calculate
\begin{align}
QA_{free}&=\sum_{k=1}^{n}F_{\mu,k}F_{\lambda,k+1}\frac{\tau_p}{\pi}\sum_{l=1}^{k}(\tau_l+\tau_p),\\
QP_{k,k+1,isi,heis}&=\frac{\pi}{8}\left(\frac{2\tau_p}{\pi}\right)^2F_{\mu,k}F_{\lambda,k+1},
\\
Q_{free}&=0.
\end{align}
We note that after summing the pulse term over $k$, we can add the two expressions together to obtain
\begin{align}
Q_{tot}=\sum_{k=1}^{n}F_{\mu,k}F_{\lambda,k+1}\frac{\tau_p}{\pi}\left(\sum_{l=1}^{k}(\tau_l+\tau_p)+\frac{\tau_p}{2}\right).
\end{align}
We note that the second sum is exactly the time at the middle of the pulse, similarly to the other Heisenberg rules. This tells us that we can cancel this the same way, by balancing the center of mass for terms of the form $F_{\mu,k}F_{\lambda,k+1}$.

\subsection{Summary of Two Qubit Commutators}
We lastly summarize the general commutators of the form $\lbrack \mathcal{O}_\alpha, \mathcal{O}_\beta \rbrack$ over a basis of two-qubit operators, $\lbrace O_\alpha \rbrace_\alpha$, to be defined shortly. To this end, it is convienent to write a generic two-qubit Hamiltonian in the form of a $4 \times 4$ matrix $\mathcal{A}$, 
\begin{align}
    H(\mathcal{A}) &= \sum_{\mu \nu=0}^3 \mathcal{A}_{\mu \nu} \, \sigma_\mu \otimes \sigma_\nu,
\end{align}
where we have defined $\sigma_\mu = \left( \bm{1}, \vec{\sigma} \right)$ as a 4-vector of Pauli operators, including the $2 \times 2$ identity matrix $\bm{1}$. It follows that a native symmetric secular Hamiltonian can be specified by the matrix
\begin{align}
    \mathcal{A} &= \left(\begin{array}{@{}c|ccc@{}}
    0 & 0 & 0 & h_2 \\ \hline
    0 & g_0 & 0 & 0 \\
    0 & 0 & g_0 & 0 \\
   h_1 & 0 & 0 & g_0+g_1 \\
  \end{array} \right),
\end{align}
parameterized by disorder fields $h_1,h_2$ and Heisenberg/Ising interactions $g_0, g_1$. Note that we have introduced horizontal and vertical bars to visually distinguish between interactions and disorder. As explained in the main-text, under global driving mapping $S_z \to F_\mu S_\mu$ the secular Hamiltonian will depend only on this column vector $\bm F$. This representation of the 2-qubit interaction will thus transform as 
\begin{align}
		\mathcal{A} \mapsto \mathcal{A}'(\bm F)&=\left(\begin{array}{@{}c|ccc@{}}
     &  & h_2 \bm{F}^T & \\ \hline
     &  &  &  \\
   	h_1 \bm{F} &  & g_0 \bm{1} + g_1 \, \bm{F}\bm{F}^T &  \\
   	 &  &  &  \\
  \end{array} \right), \\
  &= \sum_\alpha c_\alpha \, \mathcal{A}_\alpha(\bm{F}),
	\end{align}
where $\lbrace \mathcal{A}_\alpha(\bm{F}) \rbrace_\alpha$ are judicious choice of operator basis. A particular choice of operators that is convenient to summarize the commutators is the following 
\begin{align}
    \lbrace  \mathcal{A}_\alpha(\bm{F}) \rbrace_\alpha &= \left \lbrace  \mathcal{A}_{\pm}(\bm{F}) = \left(\begin{array}{@{}c|ccc@{}}
     &  & \pm \bm{F}^T &  \\ \hline
     &  &  &  \\
   	\bm{F} &  & 0 &  \\
   	 	&  &  &  \\
  \end{array}\right), \mathcal{A}_H =\left(\begin{array}{@{}c|ccc@{}}
     & & &  \\ \hline
     &  &  &  \\
   	&  & \bm{1} &  \\
   	 &  &  &  \\
  \end{array}\right),  \mathcal{A}_I(\bm{F}) = \left(\begin{array}{@{}c|ccc@{}}
     &  &  & \\ \hline
     &  &  &  \\
   	 &  & \bm{F}\bm{F}^T &  \\
   	 &  &  &  \\
  \end{array} \right) \right \rbrace.
\end{align}
The matrix commutator between the interaction picture Hamiltonians in different frames lifts to a bracket on the basis $\mathcal{C}$ matrices, yielding surprisingly simple ``selection rules" for understanding the structure behind the first order Magnus calculation. 

Before presenting the result, we define two more interactions
        \begin{align}
            \tilde{\mathcal{A}}_I(\bm F, \bm G) &=  \left(\begin{array}{@{}c|ccc@{}}
     &  &  & \\ \hline
     &  &  &  \\
   	 &  & \left(\bm{F}\bm{G}^T + \bm{G}\bm{F}^T \right)/2 &  \\
   	 &  &  &  \\
  \end{array} \right), \\
            \mathcal{A}_A(\bm F) &=  \left(\begin{array}{@{}c|ccc@{}}
     &  &  & \\ \hline
     &  &  &  \\
   	 &  & \epsilon_{ijk} F_k &  \\
   	 &  &  &  \\
  \end{array} \right),
\end{align}
where the first one contains the Ising interaction $\mathcal{A}_I(\bm F) = \tilde{\mathcal{A}}_I(\bm F,\bm F)$ as a special case, and the second one is an anti-symmetric exchange  $\mathcal{A}_A(\bm F)$.

$\lbrack$\textbf{Disorder, Disorder}$\rbrack$ $\to$ \textbf{Disorder}

\begin{itemize}
	\item $\lbrack \mathcal{A}_\sigma(\mathbf{F}),\mathcal{A}_{\sigma'}(\mathbf{G})\rbrack = 2i \,  \mathcal{A}_{\sigma \sigma'}(\mathbf{F} \times \mathbf{G}) \qquad \sigma, \sigma' \in \pm$ 
\end{itemize}

$\lbrack$\textbf{Disorder, Interaction}$\rbrack$ $\to$ \textbf{Interaction}

\begin{itemize}
	\item $\lbrack \mathcal{A}_+(\mathbf{F}), \mathcal{A}_H \rbrack =0$
	\item  $\lbrack \mathcal{A}_+(\mathbf{F}), \mathcal{A}_I(\mathbf{G}) \rbrack = 4i \, \tilde{\mathcal{A}}_I(\mathbf{G}, \mathbf{F} \times \mathbf{G})$
	\item  $\lbrack \mathcal{A}_-(\mathbf{F}), \mathcal{A}_H \rbrack = -4 i \, \mathcal{A}_A(\mathbf{F})$
	\item  $\lbrack \mathcal{A}_-(\mathbf{F}), \mathcal{A}_I(\mathbf{G}) \rbrack = 2i((F\cdot G) \mathcal{A}_A(\mathbf{G})-\mathcal{A}_A(\mathbf{F}))$
\end{itemize}

$\lbrack$\textbf{Interaction, Interaction}$\rbrack$ $\to$ \textbf{Disorder}

\begin{itemize}
	\item $\lbrack \mathcal{A}_H, \mathcal{A}_H \rbrack = 0$
	\item $\lbrack \mathcal{A}_H, \mathcal{A}_I(\mathbf{G}) \rbrack = 0$
	\item $\lbrack \mathcal{A}_I(\mathbf{F}), \mathcal{A}_I(\mathbf{G}) \rbrack = 2i \, (F \cdot G)\, \mathcal{A}_+(\mathbf{F}\times\mathbf{G})=0$ for pulse sequences built from $\pi/2,\pi$ pulses
\end{itemize}

\section{Derivation of Second-Order Decoupling Rules}
\label{supp:secondorder}
From Eq.~(\ref{eq:SecondOrderMagnus}), we have that the cancellation condition for the second-order term is given by two integrals
\begin{align*}
   & \iiint_{0<t_3<t_2<t_1<T}c_\alpha(t_1)c_\beta(t_2)c_\gamma(t_3)+\iiint_{0<t_1<t_2<t_3<T}c_\alpha(t_1)c_\beta(t_2)c_\gamma(t_3).
\end{align*}

In order to see the similarities to the previous order, we write the above integrals as follows
\begin{align*}
   & \int_0^Tdt_1c_\alpha(t_1)\left(\iint_{0<t_3<t_2<t_1}c_\beta(t_2)c_\gamma(t_3)+\iint_{t_1<t_2<t_3<T}c_\beta(t_2)c_\gamma(t_3)\right).
\end{align*}
Noting that the coefficient here is $[\Op^\alpha,[\Op^\beta,\Op^\gamma]]$, we can again sum the terms which have $\beta$ and $\gamma$ switched, as we did with first order, and derive the following expression
\begin{align}
    H^{(2)}&=\frac1{6T} ([\Op^\alpha,[\Op^\beta,\Op^\gamma]])\int_0^Tdt_1c_\alpha(t_1)\cdot\\
    &\cdot \left( \iint_{0<t_3<t_2<t_1} c_\beta(t_2)c_\gamma(t_3)-\iint_{0<t_2<t_3<t_1} c_\beta(t_2)c_\gamma(t_3)\right.+\\&+\left.\iint_{t_1<t_2<t_3<T}c_\beta(t_2)c_\gamma(t_3)-\iint_{t_1<t_3<t_2<T}c_\beta(t_2)c_\gamma(t_3)\right).
\end{align}
The inner integrals in the above expression are the first order contribution of the $\beta,\gamma$ first order term for all times before $t_1$, minus the first order contribution of the $\beta,\gamma$ first order term for all times after $t_1$. Letting 
\begin{align} c^{(1)}_{\beta, \gamma}(t_1,t_2)&= \substack{\iint\\t_1<t_a<t_b<t_2}c_{\beta}(t_a)c_{\gamma}(t_b)-\substack{\iint\\t_1<t_b<t_a<t_2}c_{\beta}(t_a)c_{\gamma}(t_b),
\end{align}
which is exactly the first order contribution between times $t_1$ and $t_2$ of the operator $[\Op^\beta,\Op^\gamma]$, we can rewrite the expression as follows
%\begin{align}    H^{(2)}&=\frac1{6T} ([\Op^\alpha,[\Op^\beta,\Op^\gamma]])\int_0^Tdt_1c_\alpha(t_1)\cdot\\&\cdot \left(c^{(1)}_{\beta,\gamma}(t_1,T)-c^{(1)}_{\gamma,\beta}(t_1,T) \right)dt_1       \end{align}
\begin{align}
    H^{(2)}&=\frac1{6T} ([\Op^\alpha,[\Op^\beta,\Op^\gamma]])\int_0^Tdt_1c_\alpha(t_1)\left(c^{(1)}_{\beta,\gamma}(0,t_1)-c^{(1)}_{\beta,\gamma}(t_1,T) \right).
\end{align}
We then write
\begin{align}
    &\int_0^Tdt_1c_\alpha(t_1) \left(c^{(1)}_{\beta,\gamma}(0,t_1)-c^{(1)}_{\beta,\gamma}(t_1,T) \right)\nonumber\\
    =&\int_0^Tdt_1c_\alpha(t_1)c^{(1)}_{\beta,\gamma}(0,t_1)-\int_0^Tdt_1c_\alpha(t_1)\left(c^{(1)}_{\beta,\gamma}(0,T)-c^{(1)}_{\beta,\gamma}(0,t_1)\right)\nonumber
    \\=& 2\int_0^Tdt_1c_\alpha(t_1)c^{(1)}_{\beta,\gamma}(0,t_1)-c^{(1)}_{\beta,\gamma}(0,T)\int_0^Tdt_1c_{\alpha}(t_1).
\end{align}

We note that this form is identical to the first order case, with $c_{\beta}(t)$ replaced by $c^{(1)}_{\beta,\gamma}(0,t)$. We will now perform the same substitution that we did for first order to convert the integral expression into a summation expression. For simplicity, we restrict our discussion to Hamiltonians involving disorder only. In this case, $c^{(1)}_{\beta,\gamma}$ does not change over a free evolution period, i.e. looking at $c^{(1)}_{\beta,\gamma}(0,t_a)$ and $c^{(1)}_{\beta,\gamma}(0,t_b)$, for $t_a,t_b$ in the same free evolution period, the contribution to the overall term is 0, as the commutator $[\Op^{\beta},\Op^{\gamma}]$ will be 0 during this time. Thus we can write the same approximation for this term as in the 1st order term
\begin{align}
    A&=\int_0^Tdt_1c_\alpha(t_1)c^{(1)}_{\beta,\gamma}(0,t_1)\\
    &=A_{free}+\sum_{k=1}^{n-1}P_{k,k+1,\alpha,\beta,\gamma}+\sum_{k=1}^{n}C^{(2)}_{\alpha,\beta,\gamma,k},
\end{align}
where
\begin{align}
    C^{(2)}_{\alpha,\beta,\gamma,k}&=\int_{t_{k-1}+\frac{\tau_{k-1}}2}^{t_{k+1}-\frac{\tau_{k+1}}2}c_{\alpha}(t_1)dt_1\int_{t_{k-1}+\frac{\tau_{k-1}}2}^{t_1}c^{(1)}_{\beta,\gamma}(0,t_2)dt_2,\\
    P_{k,k+1,\alpha,\beta,\gamma}&=\int_0^{\pi/2}\cm_{\alpha,k}(\theta_1)rd\theta_1\int_{0}^{\theta_1}\cpl^{(1)}_{\beta,\gamma,k+1}(\theta_2)rd\theta_2-\int_{0}^{\pi/2}\cpl_{\alpha,k+1}(\theta_1)rd\theta_1\int_{\theta_1}^{\pi/2}\cm^{(1)}_{\beta,\gamma,k}(\theta_2)rd\theta_2,\\
    A_{free}&=\sum_{k=1}^{n}C_{\alpha,k}\sum_{j=1}^{k-1}C_{\beta,\gamma,j}.
\end{align}
Here $C_{\beta,\gamma,j}=C_{\beta,j}\sum_{\ell=1}^{j-1}C_{\gamma,\ell}$, and the angle terms are defined as follows:
\begin{align}
\cpl^{(1)}_{\beta,\gamma,k}(\theta)&=   \iint_{0\le\theta_1\le\theta_2\le\theta}\cpl_{\beta}(\theta_1)\cpl_{\gamma}(\theta_2)-\iint_{0\le\theta_2\le\theta_1\le\theta}\cpl_{\beta}(\theta_1)\cpl_{\gamma}(\theta_2),\\
\cm^{(1)}_{\beta,\gamma,k}(\theta)&=    \iint_{0\le\theta_1\le\theta_2\le\theta}\cm_\beta(\theta_1)\cm_\gamma(\theta_2)-\iint_{0\le\theta_2\le\theta_1\le\theta}\cm_\beta(\theta_1)\cm_\gamma(\theta_2).
\end{align}
We note that this means if the two ramp up functions are proportional, i.e. $\cpl_{\gamma}(\theta)=F_{\gamma,k}\sin(\theta)$, $\cpl_{\beta}(\theta)=F_{\beta,k}\sin(\theta)$, these terms will always be 0. There are also no $q$-terms as we are restricting to a disorder Hamiltonian. Thus, the only term left is the free evolution term with the frame-lengthening correction.

We can now plug in expressions for the explicit terms in the qubit Hamiltonian to calculate the leading second-order effects. The free evolution period is much like the lower orders
\begin{align}
A_{free}=\sum_{k=1}^nF_{\mu,k}(\tau_k+\frac{4}{\pi}\tau_p)F^{\nu,\rho}_{<k},
\end{align}
where $F^{\nu,\rho}_{<k}=\sum_{l=1}^kF_{\nu,l}(\tau_l+\frac4\pi\tau_p)(F^\rho_{<l}-F^\rho_{>l})$ is the first-order contribution given by $\nu,\rho$ through time $k$. By combining $A_{free}$ with the rest of the terms, we obtain the expression in Tab.~\ref{tab:FirstOrder}
\begin{align}
2\sum_{k=1}^n F_{\mu,k}\left(\tau_k+\frac4\pi \tau_p\right)F_{<k}^{\nu,\rho} - \overline{F}^{\mu}\overline{F}^{\nu,\rho},
\end{align}
where $\overline{F}^{\nu,\rho}=\sum_{k=1}^nF_{\nu,k}(\tau_k+\frac4\pi\tau_p)F_{<k}^\rho$ is the total first-order disorder-disorder term between axes $\nu$ and $\rho$.
\end{document}